\documentclass{aa}  

\usepackage[version=4]{mhchem}
\usepackage{graphicx}
\usepackage{txfonts}
\usepackage{multirow}
\usepackage{xcolor}
\usepackage{ulem}
\usepackage[colorlinks=true,linkcolor=black,citecolor=blue,urlcolor=blue,draft=false]{hyperref}

\begin{document} 

   \title{Bayesian Analysis of Molecular Emission and Dust Continuum of Protoplanetary Disks}
   %DuCKLinG: A Tool to Fit Molecular Emission Lines and Dust Continuum of Protoplanetary Disks Simultaneously
   %Bayesian Analysing of the Molecular Emission of Protoplanetary Disks using DuCKLinG 
   \author{T. Kaeufer\inst{1}\fnmsep\inst{2}\fnmsep\inst{3}\fnmsep\inst{4}, 
           M. Min\inst{3}, 
           P. Woitke\inst{1},
           I. Kamp\inst{2}, \and 
           A. M. Arabhavi\inst{2} }
   \institute{
   Space Research Institute, Austrian Academy of Sciences, Schmiedlstrasse 6, A-8042 Graz, Austria \\ \email{till.kaeufer@oeaw.ac.at}
        \and
    Kapteyn Astronomical Institute, University of Groningen, PO Box 800, 9700 AV Groningen, The Netherlands
        \and
   SRON Netherlands Institute for Space Research, Niels Bohrweg 4, 2333CA Leiden, The Netherlands
        \and
        Institute for Theoretical Physics and Computational Physics, Graz University of Technology, Petersgasse 16, 8010 Graz, Austria
        }

   \date{Received 11 April 2024 / Accepted 7 May 2024}

% \abstract{}{}{}{}{} 
% 5 {} token are mandatory
 
  \abstract
  % context heading (optional)
   {The Mid-InfraRed Instrument (MIRI) on board the James Webb Space Telescope (JWST) probes the chemistry and dust mineralogy of the inner regions of protoplanetary disks. The observed spectra are unprecedented in their detail and reveal a rich chemistry with strong diversity between objects. This complicates interpretations which are mainly based on manual continuum subtraction and 0D slab models.}
  % aims heading (mandatory)
   {We investigate the physical conditions under which the gas emits in protoplanetary disks. Based on MIRI spectra, we apply a full Bayesian analysis that provides the posterior distributions of dust and molecular properties such as column densities and emission temperatures.}
  % methods heading (mandatory)
   {For doing so, we introduce the Dust Continuum Kit with Line emission from Gas (DuCKLinG), a python-based model describing the molecular line emission and the dust continuum of protoplanetary disks simultaneously without large computational cost. The model describes the dust continuum emission by dust models from \cite{Juhasz2009,Juhasz2010} with precomputed dust opacities. The molecular emission is based on LTE slab models, but from extended radial ranges with gradients in column densities and emission temperatures. The model is compared to observations using Bayesian analysis, with linear regression techniques to reduce the dimension of the parameter space. We benchmark this model to a complex thermo-chemical ProDiMo model of AA\,Tau and fit the MIRI spectrum of GW\,Lup. The latter allows for a comparison to the results obtained with single slab models and hand-fitted continuum by \cite{Grant2023}.}
  % results heading (mandatory)
   {We successfully decrease the computational time of the fitting method by a factor of $80$ by eliminating linear parameters like the emission areas from the Bayesian run. This approach does not significantly change the retrieved molecular parameters. Only the calculated errors on the optically thin dust masses decrease slightly.
   For an AA\,Tau ProDiMo mock observation, we find that the retrieved molecular conditions from DuCKLinG (column densities from $3\times 10^{18} \mathrm{cm^{-2}}$ to $4\times 10^{20} \mathrm{cm^{-2}}$, radial range from $0.2\,\mathrm{au}$ to $1.2\,\mathrm{au}$, and temperature range from about $200\,\mathrm{K}$ to $400\,\mathrm{K}$) fall within the true values from ProDiMo (column densities between $4\times 10^{17} \mathrm{cm^{-2}}$ to $5\times 10^{20} \mathrm{cm^{-2}}$, radial extent $0.1\,\mathrm{au}$ to $6.6\,\mathrm{au}$, and temperature range from about $120\,\mathrm{K}$ to $1000\,\mathrm{K}$). The smaller DuCKLinG ranges can be explained by the relative flux contributions of the different parts of ProDiMo.
   The parameter posterior of GW\,Lup reinforces the results found by \cite{Grant2023}. The column densities retrieved by \cite{Grant2023} fall within the retrieved ranges in this study for all examined molecules (\ce{CO2}, \ce{H2O}, \ce{HCN}, and \ce{C2H2}). Similar overlap is found for the temperatures with only the temperature range of \ce{HCN} (from $570^{+60}_{-60}\,\rm K$ to $750^{+90}_{-70}\,\rm K$) not including the previously found value ($875\,\rm K$). This discrepancy may be due to the simultaneous fitting of all molecules compared to the step-by-step fitting of the previous study. There is statistically significant evidence for radial temperature and column density gradients for \ce{H2O} and \ce{CO2} compared to the constant temperature and column density assumed in the 0D slab models. Additionally, \ce{HCN} and \ce{C2H2} emit from a small region with near constant conditions. Due to the small selected wavelength range $13.6\,\rm \mu m$ to $16.3\,\rm \mu m$, the dust properties are not well-constrained for GW\,Lup. DuCKLinG can become an important tool to analyse molecular emission and dust mineralogy of large samples based on JWST/MIRI spectra in an automated way.}
   % conclusions, optional
   {}


   \authorrunning{T.~Kaeufer et al.}
   
   \maketitle
%
%-------------------------------------------------------------------
%Bayesian

\section{Introduction}

Many of the known exoplanets are observed within a few au of their host stars. 
The chemical composition of these planets is influenced by their formation conditions in protoplanetary disks \citep{Oberg2011,Molliere2022,Khorshid2022}. This makes these regions interesting targets to examine their chemical composition that will provide the material from which planets form.

Mid-infrared observations can probe the chemical compositions in these planet-forming regions. Spectra by the Spitzer Space Telescope showed frequent detections of molecules like \ce{H2O}, \ce{OH}, \ce{HCN}, \ce{C2H2}, and \ce{CO2} \citep{Pontoppidan2010,Salyk2011}, but were limited by the spectral resolution ($R\sim600$) and sensitivity that did not allow for detections of less abundant species and weak lines of abundant species.

The medium resolution spectrograph of the Mid-InfraRed Instrument (MIRI) on board the James Webb Space Telescope (JWST) improved the achievable spectral resolution ($R\sim1500-3000$) and especially the sensitivity (by more than two orders of magnitude) compared to Spitzer.
First observations of protoplanetary disks with MIRI detected a large variety of molecules. Among the newly detected molecules are \ce{C4H2}, \ce{C6H6} \citep{Tabone2023} \ce{CH4}, \ce{C2H4}, \ce{C3H4}, \ce{C2H6}, \citep{Arabhavi2024}, and even isotopologues like \ce{^{13}CO2} \citep{Grant2023} and \ce{^{13}C^{12}CH2} \citep{Tabone2023}.

While single efforts have been made to use complex thermo-chemical disk models to describe these observations \citep{Woitke2023}, most often simple 0D slab models are used.
These often assume local thermal equilibrium (LTE) and use a single temperature, column density, and emitting area for each molecule to fit the observed spectrum. They have been successfully used to identify multiple molecules \citep[e.g.][]{Grant2023,Perotti2023} and some studies use multiple slabs for the same molecule to describe their optically thin and thick emission \citep[e.g.][]{Tabone2023,Gasman2023,Arabhavi2024}. 

We know from thermo-chemical disk models \citep[e.g.\ ProDiMo,][]{Woitke2009} that molecules emit over a large range of temperatures and column densities. Abundance gradients can further complicate the picture. Additionally, the dust does coexist with the gas leading to interactions between continuum and line photons. All of this is completely neglected in 0D slab models. This questions the interpretability of extracted molecular conditions by simple 0D slab models. A comparison by \cite{Kamp2023} found that 0D slab models have trouble reproducing the molecular conditions in ProDiMo from a mock observation. There are some technical reasons like the line selection used in both models or the extraction of areas of significant emission from the thermo-chemical model, that might explain some of the differences. However, the simplifications of 0D slab model listed above justify that the retrieved results are benchmarked against more complex models.

Recent efforts to fit MIRI spectra with slab models included a step-by-step fitting approach. Since the models cannot describe the dust continuum, a continuum that is not physically motivated is subtracted, often manually, which introduces additional uncertainties to the results. This becomes especially difficult if the molecular emission is optically thick and forms a quasi-continuum that is indistinguishable from the dust continuum \citep[e.g.][]{Tabone2023,Arabhavi2024}. Typically, the fitting is done in an iterative process. After fitting one molecule or a group of molecules in a wavelength range that is dominated by its emission, the fit is subtracted from the spectrum to allow for a fit of the next molecule. This procedure accumulates the errors of each subtraction to all following iterations. Additionally, the method reaches its limits, when the molecular lines of different molecules overlap heavily as it makes disentangling the different molecular contributions difficult. 
Often grids of slab models are used to select the best model for an observation based on $\chi^2$-minimization. However, some parameters are known to be degenerate, e.g. the emitting area and column density will be completely degenerate if the emission is optically thin. A Bayesian treatment can address this by providing the full posterior for all input parameters.

The CLIcK (Continuum and Line fItting Kit) tool \citep{Liu2019} tackles some of the shortcomings of 0D slab models by describing the dust continuum and gas emission at the same time. CLIcK uses hydrostatic, passive disk models \citep{Chiang1997} with a puffed-up inner rim \citep{Dullemond2001} to describe the dust continuum, with gas emission from column density and temperature power laws added on top. This physical description of the continuum comes at the cost of computational speed, which limits the number of molecules possible to fit and the ability to run a full Bayesian analysis.

In this paper, we aim to take 0D slab models and their fitting procedure one step further by introducing a 1D advanced model called Dust Continuum Kit with Line emission from Gas (DuCKLinG). It is based on pre-calculated molecular slab models coupled with dust emission models and is fast enough to enable a full Bayesian analysis for multiple molecules. This tool interpolates in a grid of slab models to predict the emission along a radial temperature power law with varying column densities, instead of single values for both quantities. This gives the model more flexibility and a closer representation of reality. \cite{Banzatti2023b} detected excess line flux from cold water in compact disks compared to extended disks, probably tracing the sublimation of drifting pebbles and therefore a different temperature component. \cite{Gasman2023} found that the water emission in the JWST/MIRI spectrum of Sz\,98 is consistent with a radial temperature gradient. 
Similarly, \cite{Banzatti2023a} described the water emission by multiple slab models of changing conditions for objects in a sample of disks. These studies underpin the benefit of including temperature gradients for the emission from molecules in disks. The molecular emission in our model is combined with dust emission based on the models by \cite{Juhasz2009, Juhasz2010} which makes a continuum subtraction unnecessary. Since this model does not use a physical dust disk structure like CLIcK, but a simple combination of optically thick and optically thin dust components, the computational speed is massively improved. Therefore, a full Bayesian approach to extract molecular and dust parameters with their uncertainties based on likelihood calculations is possible, without the need for an iterative procedure.

This paper is structured as follows. Section \ref{sec:method} introduces the model and fitting procedure. The results are presented in Sect.~\ref{sec:results} divided into validation of the fitting procedure, a comparison to a complex thermo-chemical disk model, and the application to the JWST/MIRI observation of GW\,Lup. After discussing the need for diverse molecular conditions to fit the GW\,Lup MIRI observation in Sect.~\ref{sec:discussion}, the paper is concluded with the main finding in Sect.~\ref{sec:summary}.

\section{Method\label{sec:method}}

\subsection{The model\label{sec:model}}

\begin{figure}[t]
    \centering
    \includegraphics[width=\linewidth]{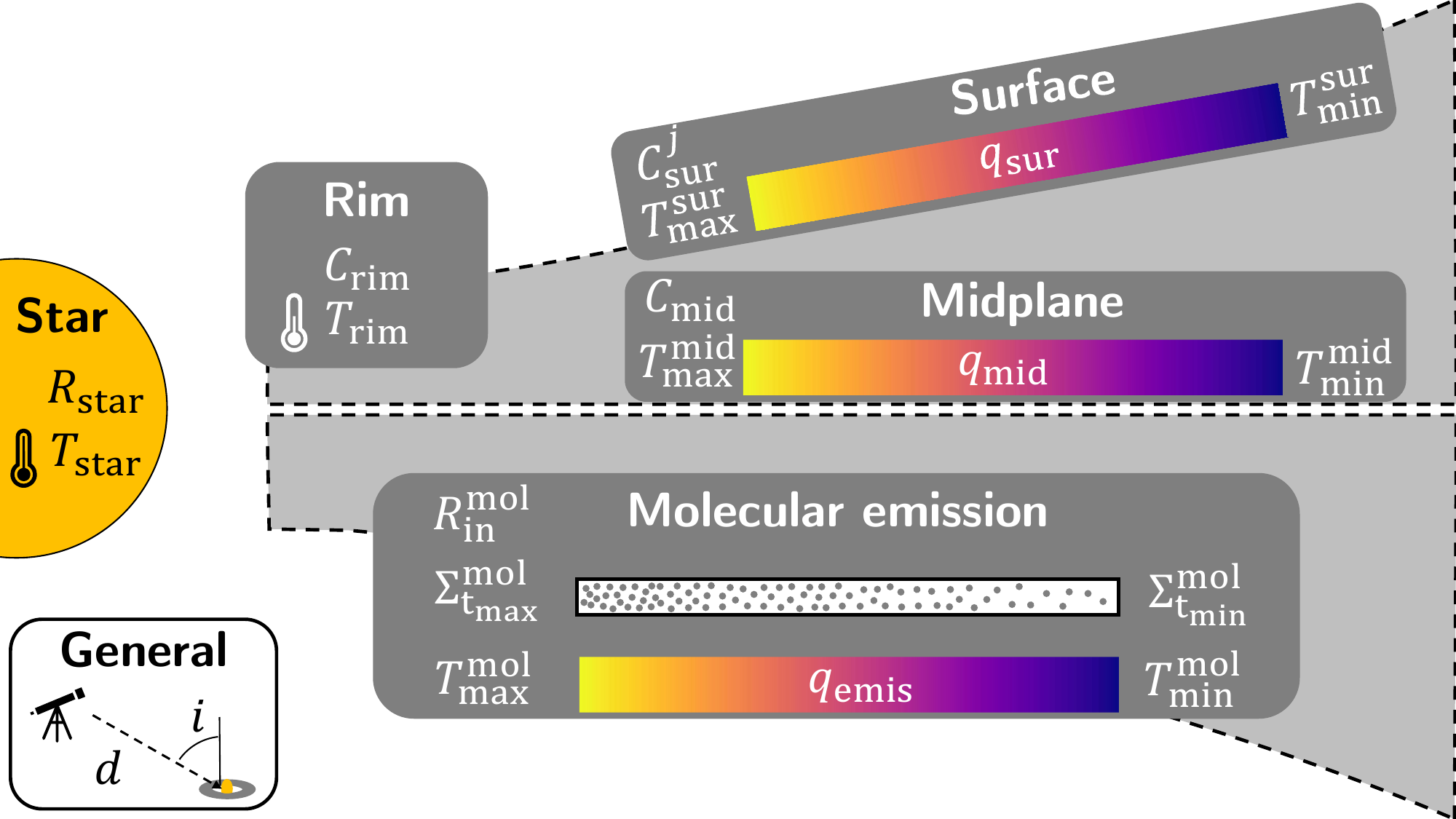}
    \caption{Sketch of the model including all free model parameters, which are further explained in Table~\ref{tab:model_parameters}.}
    \label{fig:sketch_model}
\end{figure}

The model superimposes the flux of five distinct components to derive a total flux. These components represent the star, the optically thick inner rim of the disk, the optically thick dust midplane, the optically thin dust layer, and the molecular emission layer. The first four components are based on dust models by \cite{Juhasz2009,Juhasz2010} while the molecular emission is computed using ProDiMo slab models \citep{Arabhavi2024}. 
The different components are visualized in Fig.~\ref{fig:sketch_model}.

The flux of the star ($F_\nu^{\rm star}$) is determined either through a stellar spectrum or a black body ($\mathcal{B}_{\nu}$) with a given temperature $T_{\rm star}$ using the object's distance ($d$) and the stellar radius $R_{\rm star}$:
\begin{align}
    F_\nu^{\rm star} = \frac{\pi R_{\rm star}^2}{d^2} \, \mathcal{B}_{\nu}(T_{\rm star})\ . 
\end{align}
Another black body with a temperature $T_{\rm rim}$ with its effective surface area $A_{\rm rim}$ represents the flux ($F_\nu^{\rm rim}$) arising from the inner rim of the disk: %\tk{rim power law if we use that.} 
\begin{align}
    F_\nu^{\rm\,rim} = \underbrace{\frac{A_{\rm rim}}{d^2}}_{C_{\rm rim}} \, \mathcal{B}_{\nu}(T_{\rm rim}).
\end{align}
This component characterizes the hot, optically thick dust emission typically originating from the inner rim of a protoplanetary disk. The scaling factor ($C_{\rm rim}$) as for all continuum components, is introduced because of its use during the fitting procedure (Sect.~\ref{sec:fitting}). 

The midplane component describes the optically thick dust emission originating at the disk layer where the optical depth is $1$. 
We describe the bulk this component by black bodies integrated from the inner ($R_{\rm min}^{\rm mid}$) to the outer radius ($R_{\rm max}^{\rm mid}$) of this disk component:  
\begin{align}
    F^{\rm mid}_{\nu} = \cos{i}\int\displaylimits_{R_{\rm min}^{\rm mid}}^{R_{\rm max}^{\rm mid}} \frac{2\pi r}{d^2}\mathcal{B}_{\nu}(T_{\rm mid}(r))\,dr. \label{eq:integral_mid}
\end{align}
The inclination of the disk is noted as $i$. We note that the inclination of a model is treated as a simple scaling factor of the observed area and does not account for effects like occultations and different ray paths. A radial temperature power law with the exponent $q_{\rm mid}$ is assumed as the midplane temperature profile ($ T_{\rm mid}$),
\begin{align}
    T_{\rm mid}(r) = T_{\rm max}^{\rm mid} \left( \frac{r}{R_{\rm min}^{\rm mid}}\right)^{q_{\rm mid}}, \label{eq:temp_powerlaw_mid}
\end{align}
with ($T_{\rm max}^{\rm mid}$) being the midplane temperature at the inner radius.
Translating the integral in Eq.~(\ref{eq:integral_mid}) to temperature space using Eq.~(\ref{eq:temp_powerlaw_mid}) results in 

\begin{align}
    F_\nu^{\rm\,mid} = \underbrace{\frac{-2\pi\,\cos{i}\left(R_{\rm min}^{\rm mid}\right)^2}{d^2q_{\rm mid} {T_{\rm max}^{\rm mid}}^{(2/q_{\rm mid})}}}_{C_{\rm mid}} \, \int\displaylimits_{T_{\rm min}^{\rm mid}}^{T_{\rm max}^{\rm mid}} \mathcal{B}_{\nu}(T)\, T^{(2-q_{\rm mid})/q_{\rm mid}} \,dT ,\label{eq:midplane}
\end{align}
where the scaling factor $C_{\rm mid}$ incorporates all variables outside the integral and $T_{\rm min}^{\rm mid}$ describes the midplane temperature at $R_{\rm max}^{\rm mid}$. The derivation of Eq.~(\ref{eq:midplane}) is shown in Appendix~\ref{app:model_paras}.

The total emission of the optically thin dust is described by the component that we call the surface layer.
Analogous to the midplane, we define the optically thin surface dust emission in our model as the radial integral of emission from $R_{\rm min}^{\rm sur}$ to $R_{\rm max}^{\rm sur}$, with the additional factor of the cross-section ($\sigma_{\nu}^{j}$) of every dust species ($j$) multiplied by the dust column number density $N_{\rm d}^{j}$:
\begin{align}
    F^{\rm sur}_{\nu,j} = \cos{i}\int\displaylimits_{R_{\rm min}^{\rm sur}}^{R_{\rm max}^{\rm sur}} \frac{2\pi r}{d^2} N_{d}^{j}\, \sigma_{\nu}^{j} \, \mathcal{B}_{\nu}(T_{\rm sur}(r)) \, dr. \label{eq:integral_sur}
\end{align}
We assume radius-independent column densities for all dust species with the same radial temperature power law (analogous to Eq.~(\ref{eq:temp_powerlaw_mid})) with an exponent $q_{\rm sur}$. $T_{\rm max}^{\rm sur}$ and $T_{\rm min}^{\rm sur}$ denote the temperature at the inner and outer radius, respectively. Therefore, Eq.~(\ref{eq:integral_sur}) can be transformed to
\begin{align}
    F_{\nu,j}^{\rm\,sur} = \underbrace{\frac{-2\pi\,\cos{i}\left(R_{\rm min}^{\rm sur}\right)^2}{d^2q_{\rm sur} {T_{\rm max}^{\rm sur}}^{(2/q_{\rm sur})}}\, N_{d}^{j}}_{C_{\rm sur}^{j}} \, \sigma_{\nu}^{j} \, \int\displaylimits_{T_{\rm min}^{\rm sur}}^{T_{\rm max}^{\rm sur}} \mathcal{B}_{\nu}(T)\; T^{(2-q_{\rm sur})/q_{\rm sur}} \ dT\ . \label{eq:surface}
\end{align}
The factors outside the integral, excluding the opacity, are summarised as $C_{\rm sur}^{j}$.

For the opacities of the optically thin dust component we assume single sized, homogeneous particles. For each dust material we pre-compute the opacities for a fixed number of particle sizes and treat each material and each particle size as a separate, independent dust component. The opacities are computed using the DHS method \citep{Min2005} with the irregularity parameter $f_\mathrm{max}=0.8$, simulating irregularly shaped grains. The refractive index as a function of wavelength is taken from laboratory measurements (see Table~\ref{tab:dustrefs}).

\begin{table}[!tb]
\begin{center}
\caption{References for the laboratory data for the dust species.}
\begin{tabular}{l|l}
\hline
\hline
Dust component	&	refractive index references\\
\hline
Am Mg-Olivine		&	{\cite{Jager2003}}   \\
Am Mg-Pyroxene	&	{\cite{Dorscher1995}}    \\
Silica			&	{\cite{Palik1985, Henning1997}}\\
Enstatite			&	for $\lambda>8\,\mu$m: {\cite{Jager1998}}\\
				&	for $\lambda<8\,\mu$m: {\cite{Dorscher1995}}\\
Forsterite			&	for $\lambda>2\,\mu$m: {\cite{Fabian2001}}\\
				&	for $\lambda<2\,\mu$m: {\cite{Zeidler2011}}\\
\hline
\hline
\end{tabular}
\label{tab:dustrefs}
\end{center}
\end{table}

The cumulative flux of the stellar, rim, midplane, and surface components has been demonstrated to sufficiently explain the mid-infrared flux of protoplanetary disks as observed by Spitzer \citep{Juhasz2009,Juhasz2010}.

Here we expand this dust model to encompass a gaseous component. 
The flux $F_\nu^{\rm mol}$ of each molecule ($\rm mol$) is described by an integral of specific intensities $I_{\nu}^{\rm mol}$ along the radius of the emitting region:
\begin{align}
    F_\nu^{\rm mol}=\frac{\cos{i}}{d^2}\int\displaylimits_{R_{\rm min}^{\rm mol}}^{R_{\rm max}^{\rm mol}} I_{\nu}^{\rm mol}\big(T_{\rm mol}(r),\Sigma_{\rm mol}(r)\big) \,2\pi r\,dr. \label{eq:emis_radial}
\end{align}
The temperature distribution for each molecule is a radial temperature power law with an exponent of $q_{\rm emis}$ (same value for all molecules), and the molecules emit within individual radial (from $R_{\rm min}^{\rm mol}$ to $R_{\rm max}^{\rm mol}$) and temperature ranges (from $ T_{\rm min}^{\rm mol}$ to $T_{\rm max}^{\rm mol}$):
\begin{align}
    T_{\rm mol}(r) = T_{\rm max}^{\rm mol} \left( \frac{r}{R_{\rm min}^{\rm mol}}\right)^{q_{\rm emis}}.\label{eq:temp_powerlaw_emis}
\end{align}
The column density of every molecule ($\Sigma_{\rm mol}(r)$) is defined as another radial power law between the column density $\Sigma_{\rm tmax}$ at $R_{\rm min}^{\rm mol}$/$T_{\rm max}^{\rm mol}$ and the column density $\Sigma_{\rm tmin}$ at $R_{\rm max}^{\rm mol}$/$T_{\rm min}^{\rm mol}$
\begin{align}
    \Sigma_{\rm mol}(r) = \Sigma_{\rm tmax}^{\rm mol} \left( \frac{r}{R_{\rm min}^{\rm mol}}\right)^{p_{\rm mol}} \Rightarrow 
    \Sigma_{\rm mol}(T) = \Sigma_{\rm tmax}^{\rm mol} \left( \frac{T}{T_{\rm max}^{\rm mol}}\right)^\frac{p_{\rm mol}}{q_{\rm emis}} ,\label{eq:coldens_powerlaw_emis}
\end{align}
where the exponent $p_{\rm mol}$ is defined by $\Sigma_{\rm tmax}$ and  $\Sigma_{\rm tmin}$.
The radial integral in Eq.~(\ref{eq:emis_radial}) can be substituted into a temperature integral using Eq.~(\ref{eq:temp_powerlaw_emis}) and Eq.~(\ref{eq:coldens_powerlaw_emis}), with $C_{\rm mol}$ summarizing all factors outside the integral:
\begin{align}
F_\nu^{\rm mol}= \underbrace{\frac{-2 \pi \,\cos{i} \,  \left(R_{\rm min}^{\rm mol}\right)^2}{d^2 \, q_{\rm emis} \, {\left(T_{\rm max}^{\rm mol}\right)}^{2/q_{\rm emis}}}}_{C_{\rm mol}} \, \int\displaylimits_{T_{\rm min}^{\rm mol}}^{T_{\rm max}^{\rm mol}} I_{\nu}^{\rm mol}\big(T,\Sigma_{\rm mol}(T)\big) \, T^{\frac{2- q_{\rm emis}}{q_{\rm emis}}}\,dT\ . \label{eq:emission}
\end{align}
This treatment allows for each molecule to emit over a temperature and column density range instead of single temperature and column density slabs, granting the model more flexibility.
The emission $I_{\nu}^{\rm mol}$ of each molecule at a given temperature $T$ and column density $\Sigma_{\rm mol}$ is estimated by linear interpolation in the pre-calculated grids of ProDiMo 0D slab models introduced in \cite{Arabhavi2024}. These grids span from $25\,\mathrm{K}$ to $1500\,\mathrm{K}$ and from $10^{14}\rm cm^{-2}$ to $10^{24.5}\rm cm^{-2}$, in steps of $25\,\mathrm{K}$ temperature and $1/6\,\rm dex$ column density, respectively. The model spectra cover the MIRI wavelength range with a very high spectral resolution of $R=10^5$ to account for line opacity overlap which is important for molecules that are extremely optically thick. The resulting spectra are convolved with $R=3000$. We note that for this particular project, the narrow fitted wavelength region of GW\,Lup justifies the use of a constant $R$. Changing $R$ to $2500$ does not significantly impact the retrieved parameter values. The level populations are calculated in local thermal equilibrium (LTE) and the line profiles are assumed to be Gaussian. The details of the slab models can be found in \cite{Arabhavi2024}.

These grids exist for \ce{H2O} (o-\ce{H2O}/p-\ce{H2O}\,$=\!3$), \ce{CO2}, \ce{HCN}, \ce{HC3N}, \ce{CH4}, \ce{C2H2}, \ce{C2H4}, \ce{C2H6}, \ce{C3H4} (only up to $600\,\mathrm{K}$), \ce{C4H2}, and \ce{C6H6} (only up to $600\,\mathrm{K}$). The spectroscopic data and partition sums are largely taken from the HITRAN 2020 database \citep{Grodon2022}. For \ce{C3H4} and \ce{C6H6}, the data are taken from \cite{Arabhavi2024}. The LAMDA database \citep{Schoier2005} is used for \ce{H2O}\footnote{We note that some very high excitation water lines that are visible in the fitted GW\,Lup spectrum (at about $15.2\,\rm \mu m$) are missing in this data collection.}. Molecules for which the spectroscopic data of the isotopologues are available, isotopic abundance fractions of $70$ and $35$ are used for one and two carbon-bearing hydrocarbons respectively \citep{Woods2009}.

%The column density is set at the minimum and maximum temperature of the emitting molecule and follows an exponential function in between. Alternatively, the column density can be fixed to a single value for all temperatures. 

\begin{figure}
    \centering
    \includegraphics[width=\linewidth]{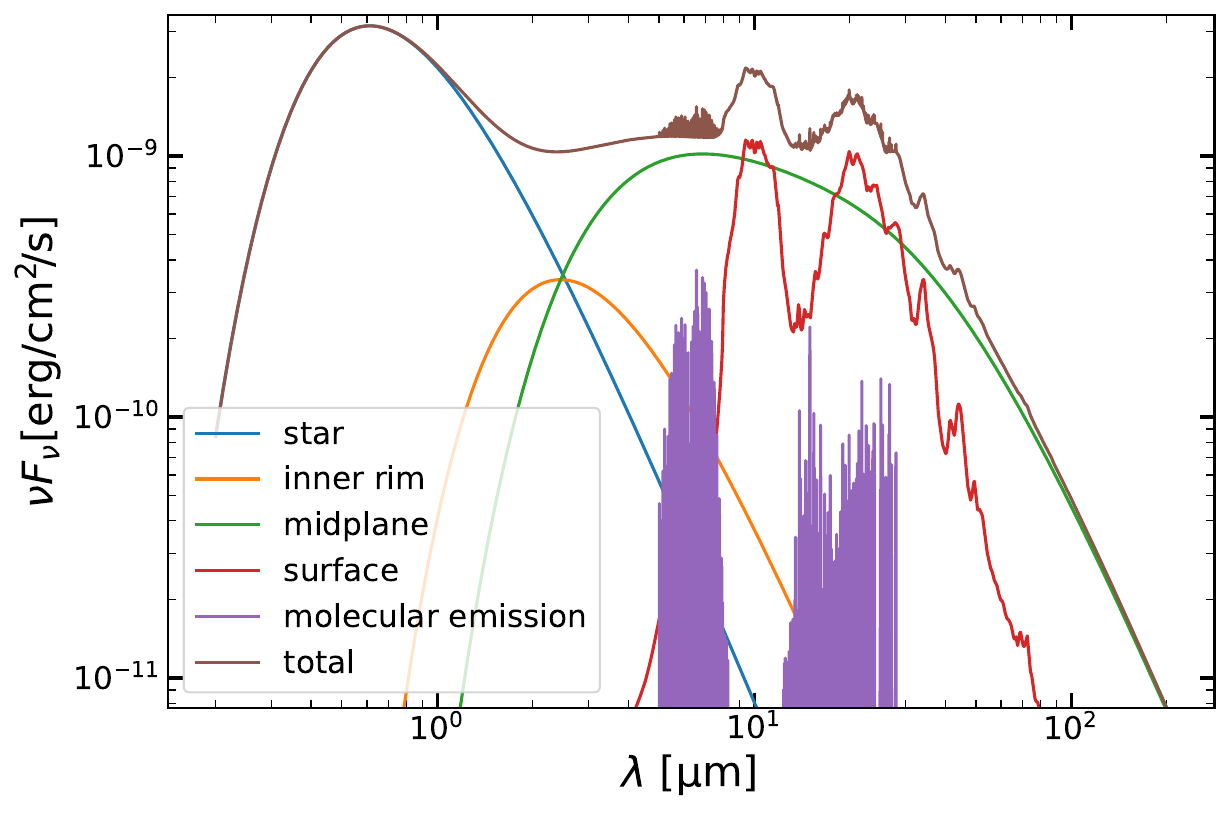}
    \caption{Example model. The brown line shows the total flux of the model. This flux is a combination of stellar flux (blue), rim flux (orange), midplane flux (green), surface flux (red), and molecular emission (purple).}
    \label{fig:toy_model}
\end{figure}

\begin{table}[ht]
    \centering
    \caption{Free parameters of the model}
    \label{tab:model_parameters}
    \vspace*{-3mm}
    \resizebox{78mm}{!}{
    \begin{tabular}{l l} \hline \hline
          & \\[-1.9ex]
          symbol &  explanation  \\ \hline
          & \\[-1.9ex]
          & general parameters \\
          & \\[-1.9ex]
          $d \, [\mathrm{pc}]$ & distance \\ 
          & \\[-1.9ex]
          $i \, [\mathrm{^\circ}]$ & inclination \\ \hline
          & \\[-1.9ex]
          & stellar parameters\tablefootmark{(1)} \\
          & \\[-1.9ex]
          $T_{\rm star} \, [\mathrm{K}]$& temperature \\
          & \\[-1.9ex]
          $R_{\rm star} \, [\mathrm{R_\odot}]$& radius \\ \hline
          & \\[-1.9ex]
          & disk rim parameters \\
          & \\[-1.9ex]
          $T_{\rm rim} \, [\mathrm{K}]$& temperature \\
          & \\[-1.9ex]
          $C_{\rm rim}$ & scaling factor \\ \hline
          & \\[-1.9ex]
          & disk midplane parameters \\
          & \\[-1.9ex]
          $T_{\rm min}^{\rm mid} \, [\mathrm{K}]$ & minimum temperature\\
          & \\[-1.9ex]
          $T_{\rm max}^{\rm mid} \, [\mathrm{K}]$ & maximum temperature \\
          & \\[-1.9ex]
          $q_{\rm mid}$ & exponent of the temperature distribution \\
          & \\[-1.9ex]
          $C_{\rm mid}$ & scaling factor\\ \hline
          & \\[-1.9ex]
          & disk surface layer parameters \\
          & \\[-1.9ex]
          $T_{\rm min}^{\rm sur} \, [\mathrm{K}]$ & minimum temperature\\
          & \\[-1.9ex]
          $T_{\rm max}^{\rm sur} \, [\mathrm{K}]$ & maximum temperature \\
          & \\[-1.9ex]
          $q_{\rm sur}$ & exponent of the temperature distribution \\         
          & \\[-1.9ex]
          $C_{\rm sur}^{j}$ & abundance factor of a dust species\tablefootmark{(2)}\\\hline
          & \\[-1.9ex]
          & molecular emission parameters \\
          & \\[-1.9ex]
          $q_{\rm emis}$ & exponent of the temperature distribution \\
          & \\[-1.9ex]
          $\Sigma_{\rm tmin}^{\rm mol} \, [\mathrm{cm^{-2}}]$ & column density at $T_{\rm min}^{\rm mol} $ of a molecule\tablefootmark{(3)} \\
          & \\[-1.9ex]
          $\Sigma_{\rm tmax}^{\rm mol} \, [\mathrm{cm^{-2}}]$ & column density at $T_{\rm max}^{\rm mol} $ of a molecule\tablefootmark{(3)} \\
          & \\[-1.9ex]
          $T_{\rm min}^{\rm mol} \, [\mathrm{K}]$ & minimum temperature of a molecule\tablefootmark{(3)} \\
          & \\[-1.9ex]
          $T_{\rm max}^{\rm mol} \, [\mathrm{K}]$ & maximum temperature of a molecule\tablefootmark{(3)} \\
          & \\[-1.9ex]
          $R_{\rm min}^{\rm mol} \, [\mathrm{au}]$ & inner radius of a molecule\tablefootmark{(3)} \\
    \end{tabular}}
    \tablefoot{\\
\tablefoottext{1}{Alternatively, a stellar spectrum can be read in.}\\
\tablefoottext{2}{For each dust species.}\\
\tablefoottext{3}{For each molecule.}\\
%\tablefoottext{4}{Alternatively, a fixed column density at $T_{\rm min}$ and $T_{\rm max}$ can be assumed.}
}
\end{table}

The total model flux ($F_{\nu}$) is given by the sum of the stellar, dust continuum and molecular emission line contributions:
\begin{align}
    F_{\nu} =F_\nu^{\rm star} + F_\nu^{\rm\,rim} + F_\nu^{\rm\,mid} + \sum\limits_{j}F_{\nu,j}^{\rm\,sur} + \sum\limits_{\rm mol} F_\nu^{\rm\,mol} \ .
    \label{eq:Ftot}
\end{align}
An example of such a model is provided in Fig. \ref{fig:toy_model}. The individual components are visualized, with their summation making up the full model.

All free parameters describing the model are listed in Table~\ref{tab:model_parameters} and visualized in Fig.~\ref{fig:sketch_model}. There are $14$ free parameters plus $1$ and $5$ additional free parameters per dust species and molecule, respectively. We note that $C_{\rm rim}$ and $C_{\rm mid}$ are proportional to the physical parameters of $A_{\rm rim}$ and $R_{\rm mid}^{\rm min}$. 

The abundance factor of each dust species $C_{\rm sur}^{j}$ is proportional to their dust masses (Appendix~\ref{app:dust_mass}). Knowing the radial structure, the radius of the effective emission area ($R_{\rm eff}$) for an observer, can be calculated using the inclination ($i$), and the inner ($R_{\rm min}$) and outer radius ($R_{\rm max}$) of the component:
\begin{align}
    R_{\rm eff}^2= \cos{i} \, \left( {R_{\rm max}}^2 -{R_{\rm min}}^2\right).
\end{align}
This radius is equivalent to the emitting radius that is fitted using slab models (e.g. \citealp{Grant2023,Perotti2023}).

%This model can describe both the dust continuum and the gas emissions at the same time, thereby eliminating the need for a manual continuum subtraction.

The model can also be used without the gaseous component (DuCK: Dust Continuum Kit \sout{with Line emission from Gas}), making it a tool to examine dust compositions. Jang et al. (in prep) will present an analysis the dust composition of PDS\,70 based on JWST/MIRI observations based on DuCK.

A single DuCKLinG model describing the dust and gas simultaneously takes typically not longer than a few milliseconds to run, which allows for incorporating many molecules in a full Bayesian fitting procedure.

\subsection{Fitting procedure\label{sec:fitting}}

We fit the model to observations using MultiNest \citep{Feroz2008,Feroz2009,Feroz2019} through the Python package PyMultinest \citep{Buchner2014} to retrieve values and uncertainties for our model parameters. This Bayesian inference tool determines the posterior distribution for all parameters using multimodal nested sampling. The details of the algorithm are given in the references above. 

%This is to our knowledge the first attempt to compare JWST/MIRI observations of protoplanetary disks to a model in a Bayesian way.

The quality of a model fit for a particular observation is quantified by a likelihood function $\mathcal{L}$ that evaluates the differences between the model spectrum $F_{\rm i,\rm model}$ and the observations $F_{\rm i,\rm obs}$ with uncertainties $\sigma_{\rm i,\rm obs}$, where $i\!=\!1\,...\,N_{\rm obs}$ is an index running through a selected wavelength range:

\begin{align}
    \mathcal{L} = \prod_{i=1}^{N_{\rm obs}}
    \frac{1}{\sqrt{2\pi\, \sigma_{i,\rm obs}^2}}
    \exp{\left(-\frac{\left(F_{i,\rm model} 
    -F_{i,\rm obs}\right)^2}
    {2\,\sigma_{i,\rm obs}^2}\right)} \ .
    \label{eq:likelihood_classic}
\end{align}
We treat the observational uncertainty as a free parameter to avoid the need to manually determine the uncertainty based on the noise of a line-free region. Assuming that the uncertainty is proportional to the flux of the observation, the parameter $a_{\rm obs}$ defines $\sigma_{\rm i,obs}$ as
\begin{align}
    \sigma_{i,\rm obs}= a_{\rm obs}\, F_{i,\rm obs}. \label{eq:flux_uncertainty}
\end{align}
The model flux $F_{\rm i,\rm model}$ is calculated from Eq.\,(\ref{eq:Ftot}) at the wavelength points of the observations, using the Python package SpectRes \citep{Carnall2017} to rebin the data. This package evaluates the fraction of every model wavelength bin that falls in each observational bin in a highly optimized way. Based on this, the fluxes corresponding to the new wavelength grid are calculated. 

Likelihood calculations including the flux prediction of one model are done within milliseconds. Nevertheless, the determination of a posterior is computationally expensive due to the large number of parameters (see Table~\ref{tab:model_parameters}). A reduction of the number of parameters sampled with MultiNest is needed. This is typically done by fixing parameters to known values. The only parameters in our model with literature values are the stellar parameters, the inclination, and distance. Fixing them still leaves many free parameters (for a fit with $10$ dust species and $5$ molecules, $46$ parameters are needed).

\begin{figure}[!b]
    \centering
    \includegraphics[width=0.9\linewidth]{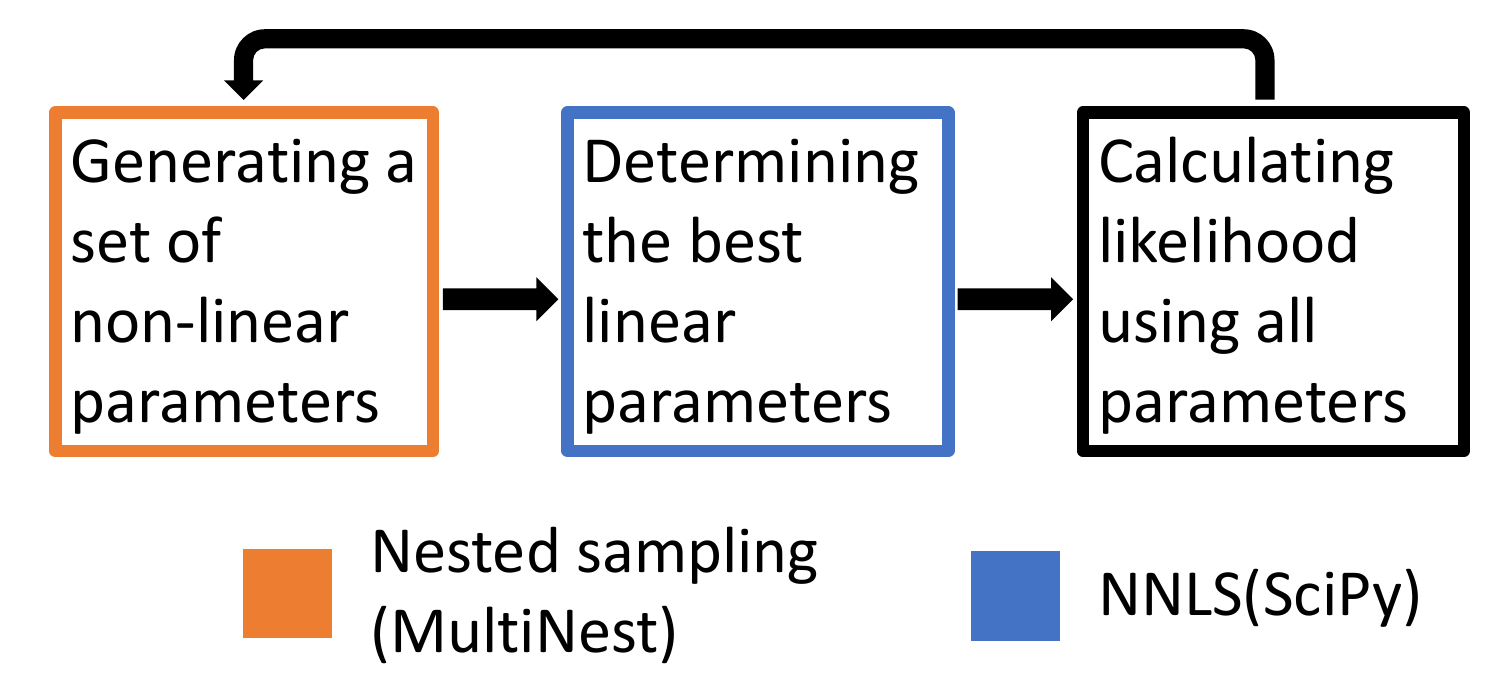}
    \caption{Sketch of the fitting procedure. While the nested sampling algorithm (orange) creates all non-linear parameters, the linear parameters (two parameters plus one per dust species and one per molecule) are determined by non-negative least squares (NNLS; blue). The total set of parameters is then used to calculate the likelihood.}
    \label{fig:sketch_linear}
\end{figure}

We decrease the parameter space further by exploiting the linearity of certain parameters.
The scaling factors ($C_{\rm rim}$, $C_{\rm mid}$), the abundances of the dust species ($C_{\rm sur}^j$), and the scaling factor of every molecule ($C_{\rm mol}$) are linear parameters. This means that they scale the respective model components without changing the shape of them. Therefore, we can use a simple non-negative least square solver (NNLS) by SciPy \citep{scipy2020} to obtain the optimal values of the linear parameters for a given set of the non-linear parameters. This leads to the procedure outlined in Fig.~\ref{fig:sketch_linear}. The likelihood of a model is calculated given its non-linear parameters. The linear parameters are calculated using NNLS, and all parameters are finally used to predict the model flux and calculate the corresponding likelihood. For the aforementioned example, this reduces the number of free parameters from $46$ to $31$.

Our fast approach comes at the cost of a less accurate Bayesian sampling of the full parameter space. We will examine the consequences of this approach in detail in Sect.~\ref{sec:validate_fitting}, where
we compare our results to those obtained using a full Bayesian sampling of all parameters.

\section{Results\label{sec:results}}

\subsection{Validating the fitting procedures}
\label{sec:validate_fitting}

We validate our procedures by fitting a DuCKLinG mock observation, to check whether we can retrieve the chosen parameter values.  We will do this twice, first with the full Bayesian analysis for all parameters, and then for the accelerated method described in Sect.~\ref{sec:fitting}.
A mock observation is created using DuCKLinG with $5$ dust species (all species listed in Table~\ref{tab:dustrefs} with a size of $2\,\mu m$ are included with identical scaling factors $C_{\rm sur}^j$) and emission from \ce{H2O}  and \ce{CO2} (both with $R_{\rm eff}=0.1\, \rm au$), with a spectral resolution of $3000$, spanning the entire JWST/MIRI wavelength range, assuming a relative flux uncertainty of $0.1\%$ for the unscattered mock observation.

For this exercise, we fix most parameters to their true values, except for the exponents of the temperature power laws ($q_{\rm mid}$, $q_{\rm sur}$, and $q_{\rm emis}$), $C_{\rm rim}$, $C_{\rm mid}$, the dust abundances, and the molecular parameters (18 parameters). The fast retrieval finds the linear parameters ($C_{\rm rim}$, $C_{\rm mid}$, all abundances of the dust species, and the emitting radii of the molecules) with NNLS, and only uses MultiNest to determine the remaining 11 parameters. The priors of the fast approach are listed in Table~\ref{tab:prior_sample_all}. The prior ranges include the true values ($q_{\rm mid}=-0.6$, $q_{\rm sur}=-0.55$, and $q_{\rm emis}=-0.55$; \ce{CO2} emitting from $200\,\rm K$/$10^{17}\,\rm cm^{-2}$ to $500\,\rm K$/$10^{21}\,\rm cm^{-2}$, \ce{H2O} emitting from $200\,\rm K$/$10^{17}\,\rm cm^{-2}$ to $900\,\rm K$/$10^{19}\,\rm cm^{-2}$). All prior ranges are chosen to be relatively narrow to increase the computational speed of the retrieval. The full approach uses additional priors for the 
$R_{\rm min}$ ($\mathcal{J}(10^{-3},10^{0})$) of both molecules and for all scaling factors (including dust abundances). Since the scaling factors can span orders of magnitude a trick is used to create narrower priors for these parameters. The scaling factor priors are defined with respect to the relative contribution (peak component flux to peak observed flux) of their components ($\mathcal{J}(10^{-2},10^{0})$). A value of $10^{-2}$ corresponds to the parameter's component having a peak flux that is $1\,\%$ of the maximum observed flux. Since all components contribute significantly more than $1\,\%$, these priors incorporate the true values for $C_{\rm rim}$, $C_{\rm mid}$, and all abundances of the dust species.

\begin{table}[t]

    \caption{Prior distributions of free parameters used for the comparison of the fast and full fitting approach. $\mathcal{U}(x,y)$ and $\mathcal{J}(x,y)$ denote uniform and log-uniform priors in the range from $x$ to $y$, respectively. The units of the parameters are given in Table~\ref{tab:model_parameters}.}
    \label{tab:prior_sample_all}
    \centering
    \begin{tabular}{l|l|l|l}
\hline \hline & & & \\[-1.9ex] 
Parameter & Prior & Parameter & Prior \\ \hline  
  & & & \\[-1.9ex] 
$q_{\rm mid}$ & $\mathcal{U}(-1,-0.1)$ &$\Sigma_{\rm tmax}^{\rm mol}$\tablefootmark{(1)} & $\mathcal{J}(10^{17},10^{22})$ \\ 
  & & & \\[-1.9ex] 
$q_{\rm sur}$ & $\mathcal{U}(-1,-0.1)$ &$T_{\rm max}^{\rm mol}$\tablefootmark{(1)} & $\mathcal{U}(200,1100)$ \\  
 & & & \\[-1.9ex] 
$q_{\rm emis}$ & $\mathcal{U}(-1,-0.1)$ &$T_{\rm min}^{\rm mol}$\tablefootmark{(1)} & $\mathcal{U}(100,500)$ \\  
 & & & \\[-1.9ex] 
$\Sigma_{\rm tmin}^{\rm mol}$\tablefootmark{(1)} & $\mathcal{J}(10^{15},10^{19})$ &  & \\ \hline
    \end{tabular}
        \tablefoot{\\
\tablefoottext{1}{Same prior used for \ce{CO2} and \ce{H2O}.}}
\end{table}

\begin{figure}[t]
    \centering
    \hspace*{-1mm}
    \includegraphics[width=1.01\linewidth]{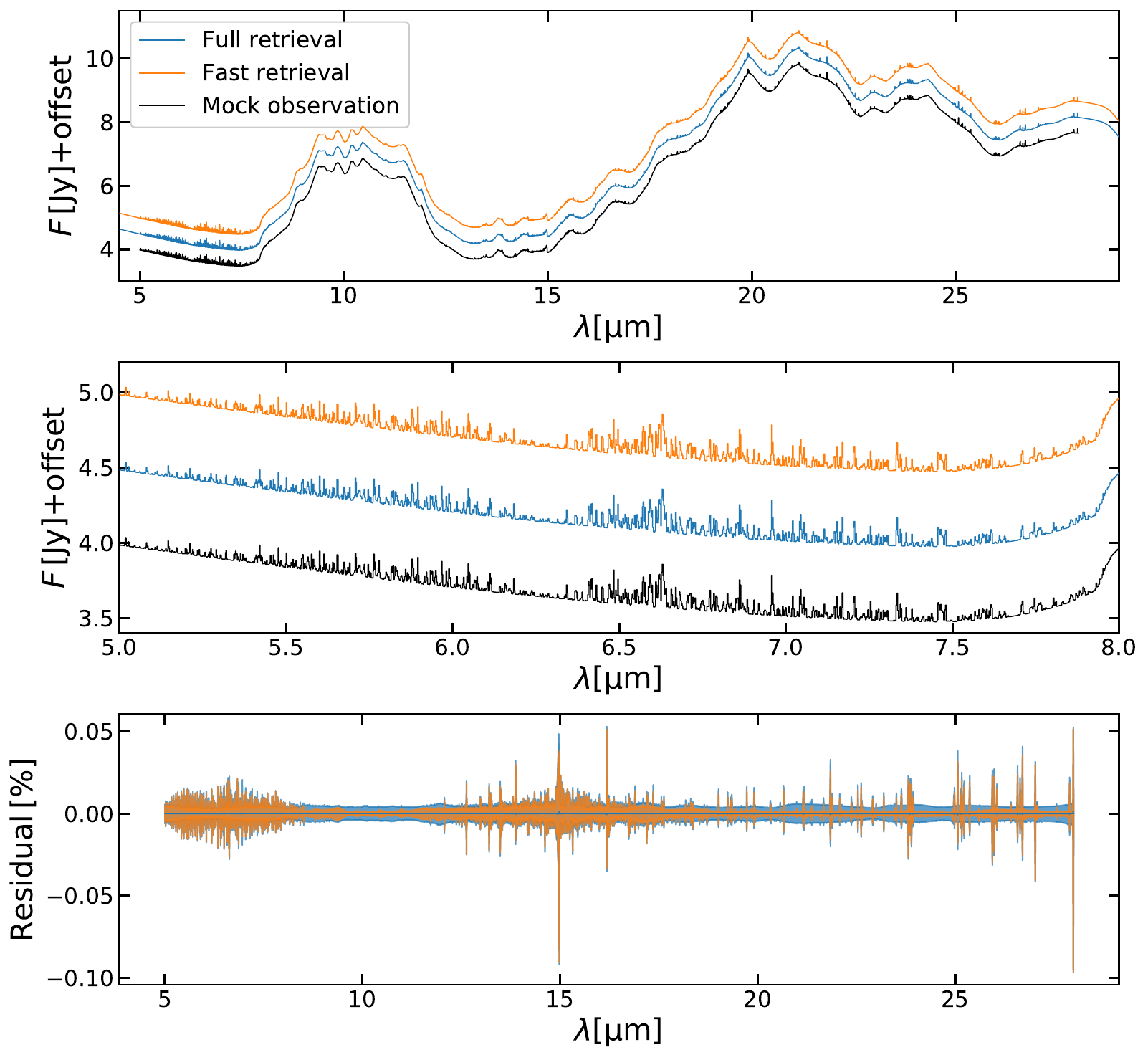}
    \caption{Comparison of the mock observations and the spectra from the two retrieval approaches. The mock observation is shown in black. The median spectrum from the posterior derived using the fast and full approach are shown in orange (shifted up by $1\,\rm Jy$) and blue (shifted up by $0.5\,\rm Jy$), respectively. While the upper panel shows the full wavelength range of the mock observation, the middle panel shows a zoom-in to the line-rich region around $6.5\,\rm \mu m$. The lower panel displays the residual (defined as ($(F_{\rm obs}-F_{\rm model})/F_{\rm obs}$) in percent between the two retrievals and the mock observation, with the coloured areas indicating the $1\sigma$ uncertainty of the retrieved fluxes.}
    \label{fig:mock_obs_spectra}
\end{figure}

Figure~\ref{fig:mock_obs_spectra} shows the mock observation compared to the median spectra from both posteriors. The retrieved spectra are shifted to allow for a visual comparison. As seen in the upper panel of Fig.~\ref{fig:mock_obs_spectra}, both retrieved spectra follow the overall shape of the mock observation well. The lower panel shows a zoom-in to the line-rich region around $6.5\,\rm \mu m$, which gives an impression of the retrieved quality of the molecular emission. The mean differences between the posterior of spectra and the mock observation are $0.0027\,\%$ and $0.0043\,\%$ for the fast and full approach, respectively (see lower panel of Fig.~\ref{fig:mock_obs_spectra}). This is well below the observational uncertainty of $0.1\,\%$. Therefore, we conclude that both fitting approaches converge towards the optimal solution, with the fast approach resulting in slightly better model spectra. We comment on this when discussing the retrieved parameters.

\begin{figure*}[ht]
    \centering
    \includegraphics[width=0.9\linewidth]{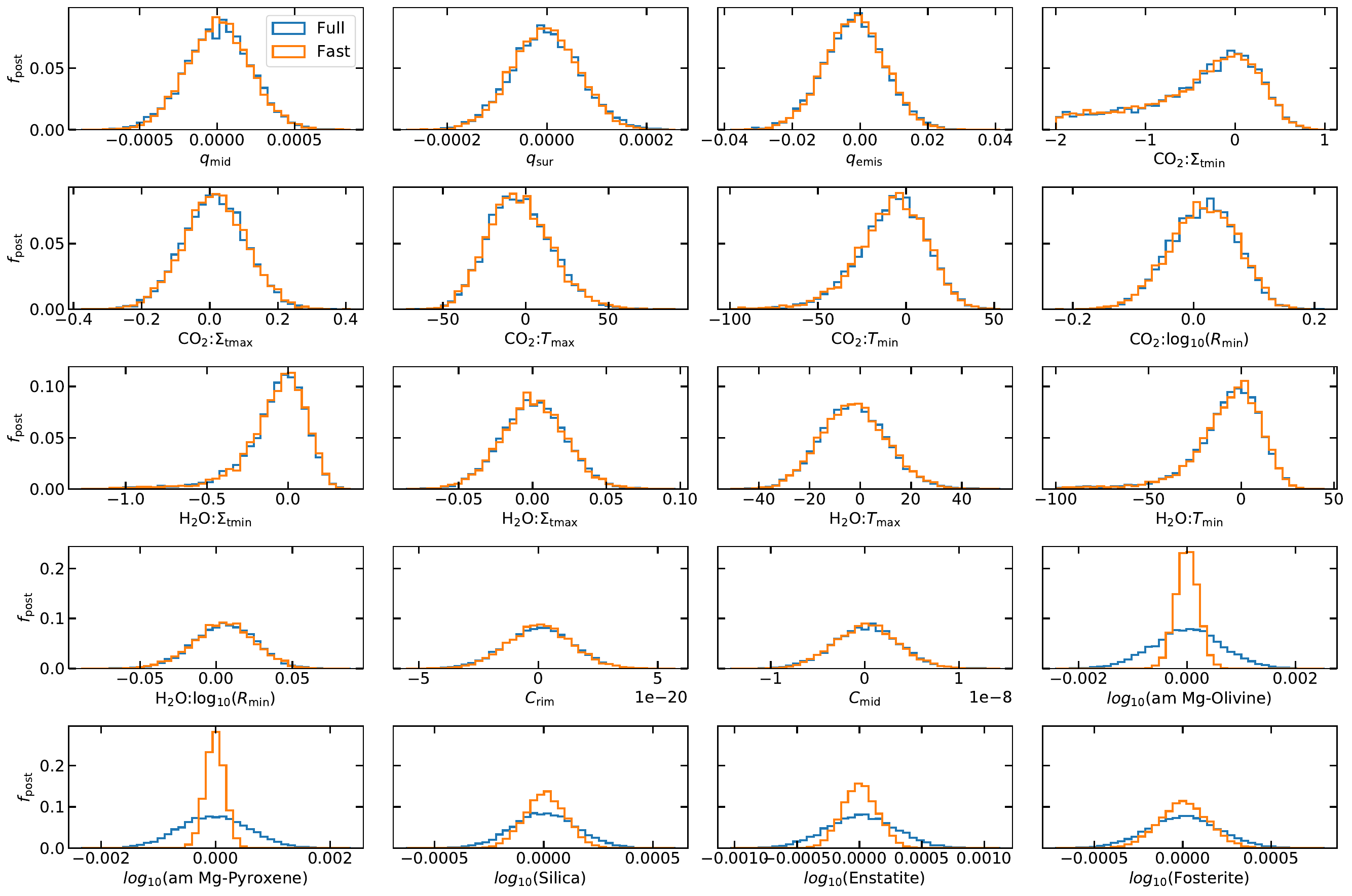}
    \caption{Histograms of the posterior of all parameters for fits of the mock observations. The results obtained by fitting all parameters with MultiNest (full) are shown with blue lines. The results obtained with the accelerated method (fast) are shown with orange lines. The horizontal axis shows the parameter differences to the values used to create the mock observation. The vertical axis displays the fraction ($f_{\rm post}$) of posterior models in the displayed bins.}
    \label{fig:sample_all}
\end{figure*}

Figure \ref{fig:sample_all} shows the posteriors for all parameters derived with the two methods. The horizontal axes show the differences between the denoted parameters and the values that were used to create the mock observation. The vertical axis displays the number of posterior models within the parameter bins. 
The posterior distributions peak for both approaches at the values used to create the mock observation (x-value of $0$ in all histograms of Fig.~\ref{fig:sample_all}). While the histograms for both approaches match perfectly for most parameters, the dust abundances show some differences. Both posteriors peak around the true value, but the fast approach consistently finds smaller uncertainties for the dust abundances. Therefore, the fast approach has an unjustified confidence in the retrieved values for these specific parameters but overall retrieved the same values. This overconfidence can be explained by an unresolved degeneracy between the abundance of Mg-Olivine and Mg-Pyroxene. While the fast approach is solving for these parameters with NNLS, and therefore jumping to the optimal solution all the time, the full approach explores these parameters in a Bayesian way and therefore accounts for the non-optimal solutions that make up the degeneracy. This also explains the slightly better fitting quality of the fast approach. However, this non-resolving of degeneracies is not seen between non-linear parameters and the other linear parameters ($\log R_{\rm min}$ for \ce{CO2}, $\log R_{\rm min}$ for \ce{H2O}, $C_{\rm rim}$ and $C_{\rm mid}$). These are resolved well with the fast method, which can be seen by the similar posterior width for all parameters except the dust species. Therefore, we conclude that the fast approach is a good representation of a full Bayesian approach, with only the uncertainties of specific parameters varying and the parameters of interest, the molecular parameters, basically unaffected.

This slightly lower accuracy comes with a huge benefit in computational speed. The full retrieval method needed about $165\,\rm million$ evaluations of the likelihood function to converge (defined by a log-evidence tolerance of $0.5$); the fast retrieval method only needed about $530\,000$ evaluations. Determining the linear parameters with NNLS increases the time for calculating the likelihood only by about $8\,\%$. The full approach can calculate slightly more models in parallel because fewer of them are accepted and used for the next iteration of the Bayesian analysis. Even accounting for all these factors, the fast retrieval method is faster by a factor of about $80$ in wall time (fast retrieval time: $\sim 25\,\rm min$, full retrieval time: $\sim 33\,\rm h$). Therefore, the full approach is only feasible for problems with many fixed parameters (like the case presented here for validation), while the fast approach is the only practical option when fitting observational data with a handful of molecules.

\subsection{Interpretation of the retrieved parameters \label{sec:prodimo}}

%The consistency of the retrieved molecular parameters is put to a test. 
While our 1D model is physically more complete compared to 0D slab models that fit every molecule individually, one after the other, they are still way more simple than complex thermo-chemical disk models, such as ProDiMo \citep{Woitke2009}, which determine the 2D chemical abundances of all species and the gas temperature structure in the disk consistently. A detailed fit of a 2D ProDiMo thermo-chemical disk model to a MIRI spectrum is computationally very challenging due to the high computational costs of about 50\,CPU-hours per disk model. \cite{Woitke2023} have recently published such a fit for the case of EX\,Lup, which is the first and so far the only case where such a fit has been presented. To achieve this computational challenge, compromises regarding the fitting procedure compared to Bayesian analysis had to be accepted. On the other hand, complex 2D disk models can play an important role in interpreting MIRI data without the need to fit them to the spectra. They can (i) check if the conditions retrieved by simpler models are possible when accounting for the physics and chemistry in disks and in which environments they arise and (ii) they can analyse the effects of disk processes (e.g. dust evolution) and disk structures (e.g. gaps, inner cavities, rounded rims) on the observable spectra \citep[e.g.][]{Anderson2021,Greenwood2019,Vlasblom2023}. Additionally, these complex models can benchmark simpler models and check how the retrieved parameters can be interpreted \citep{Kamp2023}.

In this section, we test the physical consistency of our retrieval results by fitting the mid-IR spectrum generated by a ProDiMo disk model for AA\,Tau. We will focus on a comparison of the results describing the radial extension and physical conditions in the regions responsible for the \ce{H2O} line emission.

The ProDiMo mock observation is created for the MIRI wavelength range with $R=3000$. The spectrum includes \ce{H2O} as the only molecule. For a detailed description of the underlying ProDiMo model, see \cite{Woitke2019} and \cite{Kamp2023}. 

In our retrieval model, the stellar parameters, the disk inclination, and the distance are fixed to the values used in the mock ProDiMo model for AA\,Tau \citep{Woitke2019}: $T_{\rm star}\!=\!4260\,\mathrm{K}$, $R_{\rm star}\!=\!1.6823\,\mathrm{R_\odot}$, $i\!=\!59^\circ$, and $d\!=\!137.2\,\mathrm{pc}$. 
The dust in our retrieval model is assumed to be composed of pure grains made of either am Mg-Olivine, am Mg-Pyroxene, Silica, Forsterite, or Enstatite, all with sizes of $0.1\,\rm\mu m$, $1.5\,\rm\mu m$, $2.0\,\rm\mu m$, and $5.0\,\rm\mu m$ (as introduced in Table~\ref{tab:dustrefs}). The remaining free and non-linear parameters are listed with their prior distributions in Table~\ref{tab:prior_prodimo}. The prior ranges of the water parameters are based on the extent of the underlying slab model grid (see Sect.~\ref{sec:model}). The temperature priors of the dust components are chosen to incorporate the full posterior without parameters converging towards the edges of their prior range. We chose the priors of the temperature power law exponents to include known literature values from radial \ce{CO} temperature profiles \citep[$-0.95$ to $-0.5$,][]{Fedele2016} and typical temperature slopes extracted from a thermo-chemical Herbig disk model \citep[$-0.6$ and $-0.4$,][]{Brittain2023}.

\begin{table}[t]

    \caption{Prior distributions of free parameters used for the retrieval of the ProDiMo mock spectrum of AA\,Tau. $\mathcal{U}(x,y)$ and $\mathcal{J}(x,y)$ denote uniform and log-uniform priors in the range from $x$ to $y$, respectively. The units of the parameters are given in Table~\ref{tab:model_parameters}.}
    \label{tab:prior_prodimo}
    \centering
    \begin{tabular}{l|l|l|l}
\hline \hline & & & \\[-1.9ex] 
Parameter & Prior & Parameter & Prior \\ \hline  
  & & & \\[-1.9ex] 
$T_{\rm rim}$ & $\mathcal{U}(50,1500)$ &$q_{\rm emis}$ & $\mathcal{U}(-1,-0.1)$ \\ 
 & & & \\[-1.9ex] 
$T^{\rm sur}_{\rm min}$ & $\mathcal{U}(10,500)$ &$a_{\rm obs}$ & $\mathcal{J}(10^{-5},10^{-1})$ \\ 
 & & & \\[-1.9ex] 
$T^{\rm sur}_{\rm max}$ & $\mathcal{U}(50,2000)$ &$\Sigma_{\rm tmin}$\tablefootmark{(1)} & $\mathcal{J}(10^{14},10^{24})$ \\ 
  & & & \\[-1.9ex] 
$T^{\rm mid}_{\rm min}$ & $\mathcal{U}(10,300)$ &$\Sigma_{\rm tmax}$\tablefootmark{(1)} & $\mathcal{J}(10^{14},10^{24})$ \\ 
  & & & \\[-1.9ex] 
$T^{\rm mid}_{\rm max}$ & $\mathcal{U}(50,2000)$ &$T_{\rm max}$\tablefootmark{(1)} & $\mathcal{U}(25,1500)$ \\  
 & & & \\[-1.9ex] 
$q_{\rm mid}$ & $\mathcal{U}(-1,-0.1)$ &$T_{\rm min}$\tablefootmark{(1)} & $\mathcal{U}(25,1500)$ \\  
 & & & \\[-1.9ex] 
$q_{\rm sur}$ & $\mathcal{U}(-1,-0.1)$ & & \\
    \end{tabular}
        \tablefoot{\\
\tablefoottext{1}{This prior is used for \ce{H2O}.}}
\end{table}

\begin{figure}[t]
    \centering
    \hspace*{-1mm}
    \includegraphics[width=\linewidth]{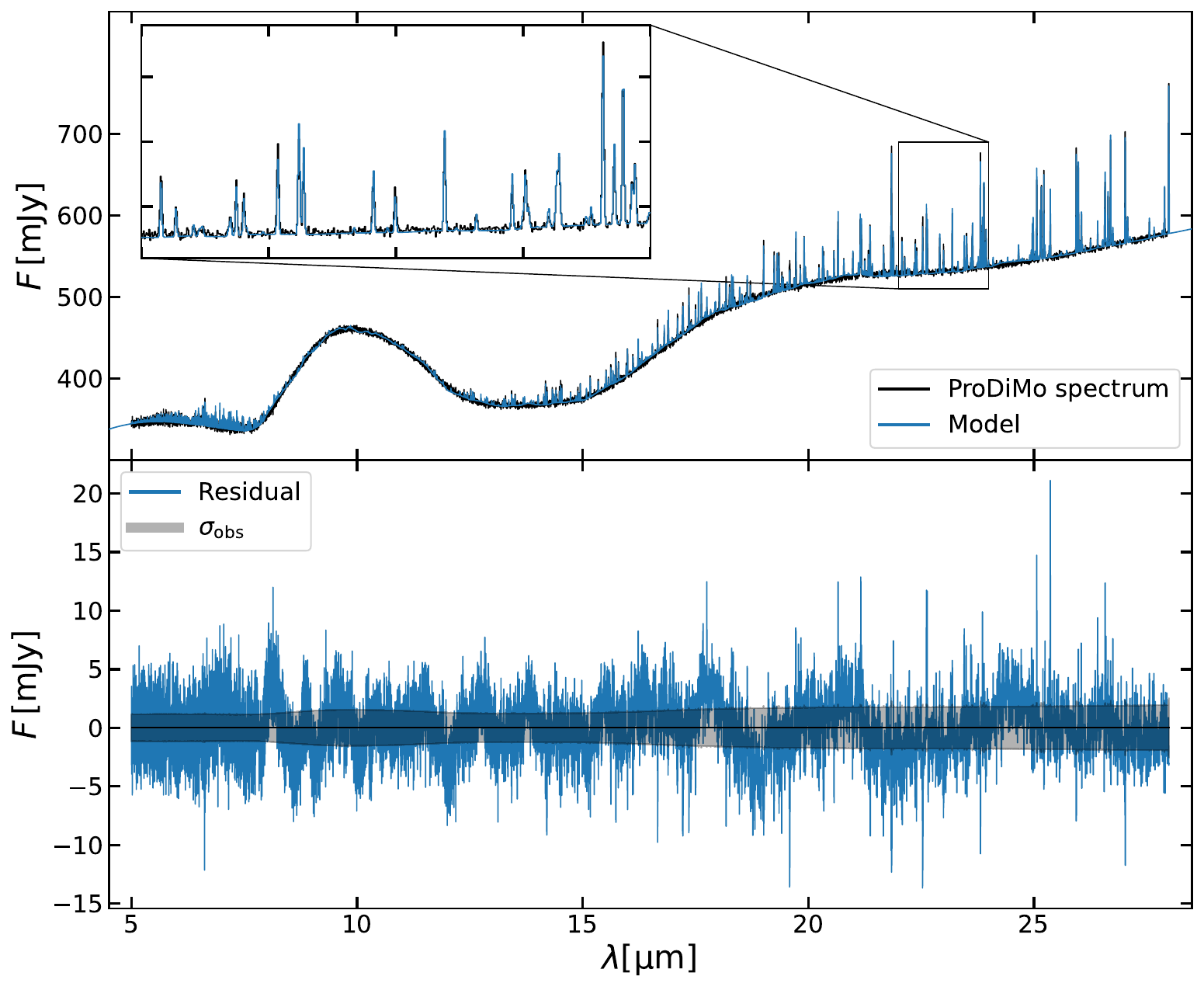}\\*[-2mm]
    \caption{Upper panel: Posterior distribution of the fit of a ProDiMo mock spectrum. The black line is the ProDiMo spectrum. The blue line shows the median fluxes from the models of the posterior distribution. The zoom-in on the upper left shows an enlarged version of the wavelength region from $22\,\rm \mu m$ to $24\,\rm \mu m$. Lower panel: Residual between the model fluxes and the ProDiMo spectrum. The blue line shows the median residual of all posterior models. The black area denotes the noise that was used to create the mock spectrum.}
    \label{fig:spectrum_prodimo}
\end{figure}

Figure~\ref{fig:spectrum_prodimo} shows that the retrieved shape of the dust features and the strengths of most water lines are reproduced well. The small deviation in the dust features (as seen in the residual in the lower panel of Fig.~\ref{fig:spectrum_prodimo}) are caused by the different dust properties used in the retrieval (Table~\ref{tab:dustrefs}) and the ProDiMo mock observation. While the $1\sigma$ deviation between model and observation ($2.5\,\rm mJy$/$0.61\,\%$) is slightly larger than the average noise ($1\sigma\!=\!1.5\,\rm mJy$) that was used to create the mock observation, more than $99\%$ of model fluxes are closer than $8\,\rm mJy$/$1.72\,\%$ to the observation. There is no large systematic deviation between model and observation (mean residual is less $0.0012\,\rm mJy$). In total, we conclude that the mock observation is well-fitted by our model. 

\begin{figure}[t]
    \centering
    \includegraphics[width=\linewidth]{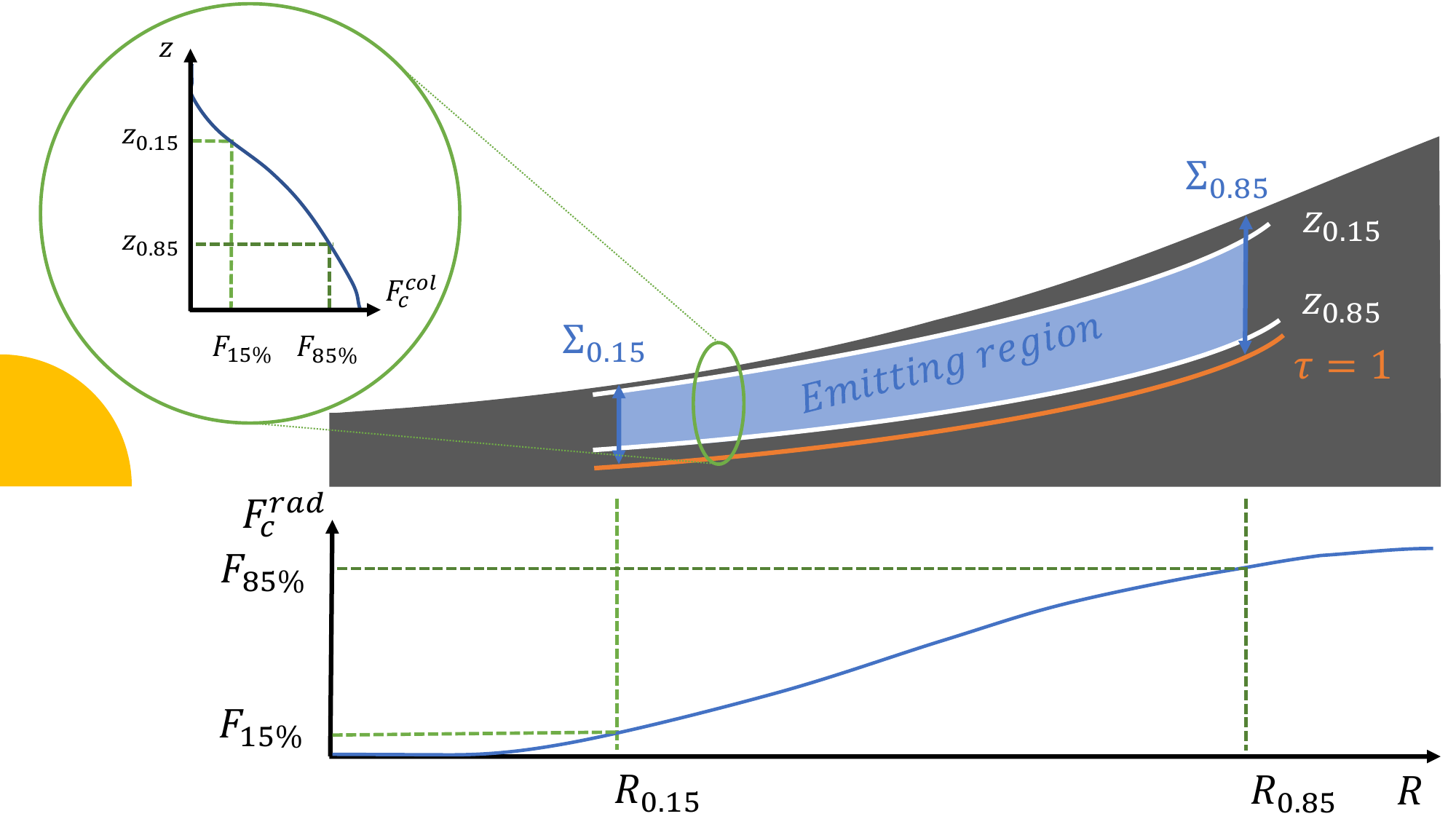}
    \caption{Sketch of the line emitting region in a ProDiMo disk model. The cumulative line flux as function of radius is shown in the lower part, and the cumulative line intensity at some given radius, as function of height, is sketched in the insert on the top left.  We consider the region between $R_{0.15}$ and $R_{0.85}$, and between $z_{0.15}(r)$ and $z_{0.85}(r)$ as the line emitting region, where the respective cumulative quantities reach $15\%$ and $85\%$, respectively.
    %of the total flux defining the inner ($R_{0.15}$) and outer ($R_{0.85}$) radius of the emitting area. In the upper left, the vertical cumulative flux is shown with the same values defining the $z_{0.15}(r)$ and $z_{0.85}(r)$ line. While the temperatures are extracted from the emitting region, the column densities are vertically defined as the column density above an optical depth of $1$ in the disk. 
    $\Sigma_{0.15}$ and $\Sigma_{0.85}$ are the column densities at $R_{0.15}$ and $R_{0.85}$, respectively.}
    %The innermost and outermost column densities in the line emitting region are $\Sigma_{\rm max}$ and $\Sigma_{\rm min}$.}
    \label{fig:sketch_prodimo}
\end{figure}

To compare the properties of the line emitting regions between the ProDiMo model and our retrieval model, we must carefully define and extract these properties from the complex 2D ProDiMo disk structure. The line emitting conditions in ProDiMo are extracted from the regions where most of the line flux originates.  Figure~\ref{fig:sketch_prodimo} illustrates this process. The radially and vertically cumulative fluxes of one specific line are used to determine the radial and vertical ranges where the central $70\,\%$ of line emission (from $R_{0.15}$ to $R_{0.85}$ and from $z_{0.15}(r)$ to $z_{0.85}(r)$ originate, respectively. This region is called the line's emitting region in ProDiMo. 
We define the emission temperature at every radius $T_{\rm emis}^{\rm line}(r)$ as the vertical mean value of the gas temperatures that we find in the line emitting region. 
Concerning the molecular column densities, we define $\Sigma_{\rm col}^{\rm line}(r)$ as the molecular column density over the height marked with $\tau\!=\!1$ in Fig.~\ref{fig:sketch_prodimo}, where the vertical continuum optical depth at line centre wavelength approaches one.
%\begin{equation}
%  \Sigma_{\rm col}^{\rm line}(r) = \int_{z\,\big(\tau_{\rm cont}^{\rm ver}(r,\lambda^{\rm
%      line})\,=1\big)}^{\infty} n_{\rm mol}(r,z)\;dz \ .
%  \label{eq:Ncol}
%\end{equation}
The column density hence depends not only on radius but also on the considered line, because the continuum optical depth depends on wavelength.  %as

%Concerning the emission temperatures for a single spectral line, we read off two mean values from the ProDiMo output, the mean gas temperature $\langle T_1\rangle$ at $R_{0.15}$ between $z_{0.15}(R_{0.15})$ and $z_{0.85}(R_{0.15})$ and the mean gas temperature $\langle T_2\rangle$ at $R_{0.85}$ between $z_{0.15}(R_{0.85})$ and $z_{0.85}(R_{0.85})$. Figure~\ref{fig:metadata_prodimo} visualises these results, connecting $(R_{0.15},\langle N_{\rm col}\rangle)$ with $(R_{0.85},\langle N_{\rm col}\rangle)$ by a horizontal grey line for a sample of water lines, and using a color-code to show  $\langle T_1\rangle$ on the left and $\langle T_2\rangle$ on the right.
%\pw{((is this correct?))} \tk{I understood from Inga that the temperatures are just the maximum and minimum in the emitting region..}

The line emitting conditions of a molecule in DuCKLinG are extracted in a similar fashion for comparability. The radially cumulative flux of \ce{H2O} determines the inner ($R_{0.15}$) and outer radius ($R_{0.85}$) of the region of significant emission (enclosing the central $70\%$ of emission). Using Eq.~(\ref{eq:temp_powerlaw_emis}) and Eq.~(\ref{eq:coldens_powerlaw_emis}) the corresponding column densities ($\Sigma_{0.15}$ and $\Sigma_{0.85}$) and temperatures ($T_{0.15}$ and $T_{0.85}$) at $R_{0.15}$ and $R_{0.85}$ are determined.
Due to the similar extraction procedure, it is expected that the extracted quantities from ProDiMo and DuCKLinG overlap if the molecular emission in both models originates under similar conditions.
Additionally, these extracted quantities from DuCKLinG are better indications of the emitting conditions of a molecule than $R_{\rm min}$/$R_{\rm max}$, $T_{\rm max}$/$T_{\rm min}$, and $\Sigma_{\rm max}$/$\Sigma_{\rm min}$. This can be seen in the example presented in Sect. \ref{sec:validate_fitting}.
The upper right panel in Fig.~\ref{fig:sample_all} shows how the retrieved posteriors of $\Sigma_{\rm min}$ for \ce{CO2}  peaks at the true value of the mock observation, but is asymmetric with a larger extension towards smaller values. This means that extending the column density power law to lower values results in similar molecular fluxes. This is underlined by the slight asymmetry of $T_{\rm min}$ and by a similar behaviour of water in Fig.~\ref{fig:sample_all}. Therefore, the lowest temperature and column density can be misleading because the emission under these conditions does not contribute significantly to the overall flux.

\begin{figure}[t]
    \centering
    \includegraphics[width=\linewidth]{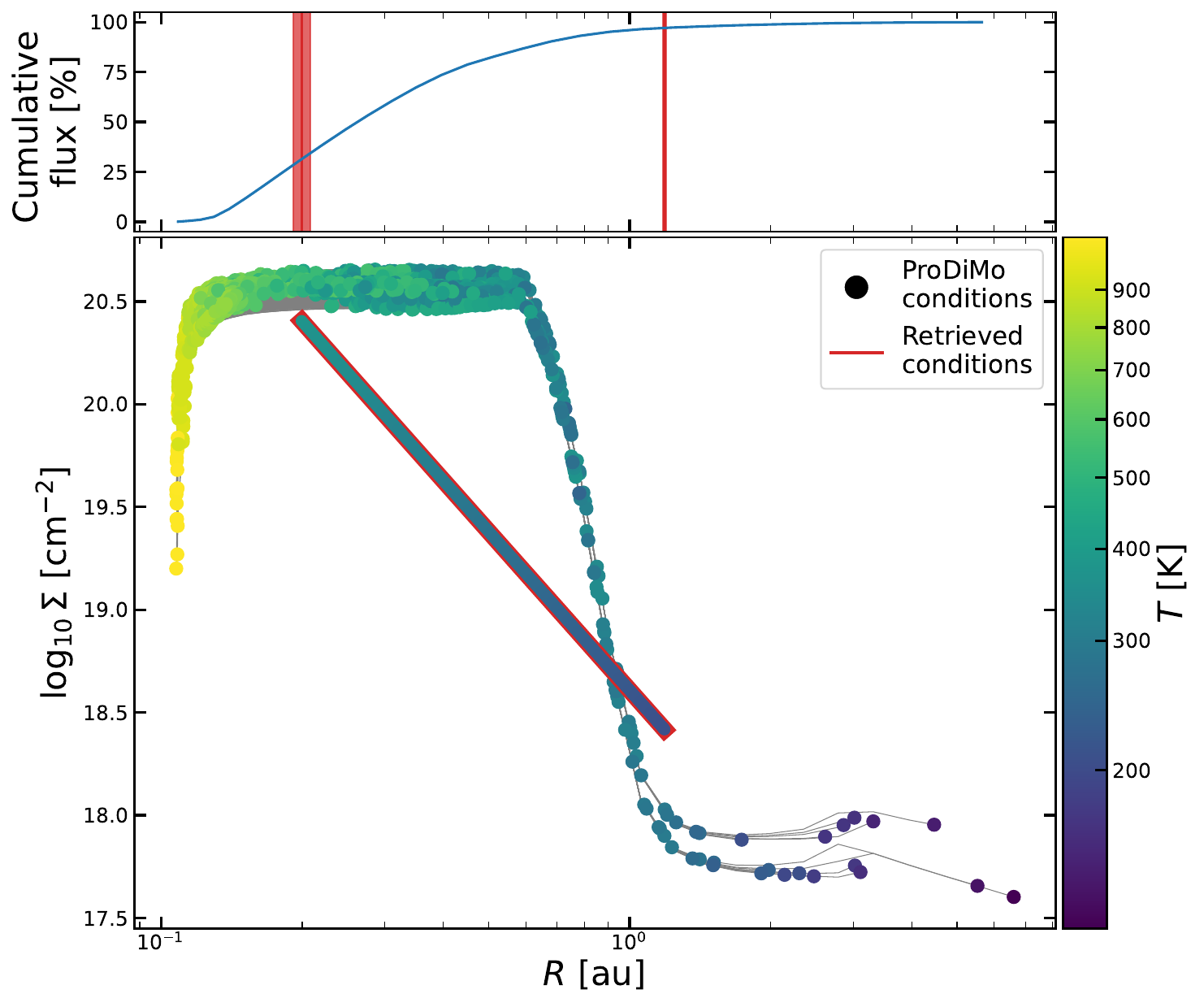}
    \caption{Comparison of the radii, column densities, and line emission temperatures of \ce{H2O} extracted from ProDiMo with the retrieved parameter conditions. Main panel: The horizontal axis shows the radius of the emission and the vertical axis the column density. Every \ce{H2O} line with peak fluxes $3\sigma$ above the observational noise is selected to show the emitting conditions in ProDiMo. The grey lines show $\Sigma_{\rm col}^{\rm line} (r)$ within the emitting region for the respective water lines. Every circle denote $R_{0.15}$ and $R_{0.85}$ for all emission lines with the color indicating the temperature $T_{\rm emis}^{\rm line}(r)$. The column density powerlaw retrieved by DuCKLinG is shown as a red outlined bar. It ranges from $R_{0.15}$ to $R_{0.85}$ with the surface colour illustrating the temperature at every radius. Upper panel: The radial cumulative flux of \ce{H2O} in ProDiMo (blue) with the from DuCKLinG retrieved $R_{0.15}$ and $R_{0.85}$ values with their uncertainties overplotted in red.}
    \label{fig:metadata_prodimo}
\end{figure}

The emission conditions of ProDiMo and the retrieved conditions by DuCKLinG are compared in Fig.~\ref{fig:metadata_prodimo}.
For ProDiMo we show $N_{\rm col}^{\rm line}(r)$ as a function of radius (grey lines) in Fig.~\ref{fig:metadata_prodimo}, using all water lines with peak line strength above the noise of the mock observation. Additionally, the column densities at $R_{0.15}$ to $R_{0.85}$ are indicated with a circle that displays the vertical mean gas temperatures at these radii colour-coded.
The retrieval results are visualised by a thick slanted bar, showing the retrieved power-law distribution of the column density as function of radius.
The colour along that bar represents the retrieved emission temperatures. 
We note that the uncertainties of the endpoints of these power laws are smaller than the displayed line width.

In the ProDiMo model, the mid-infrared water lines are emitted from the surface of the disk close to the inner rim \citep[inner part of region~3 in Fig.~1 of ][]{Woitke2009b}. The water is created by two neutral-neutral reactions, 
\begin{equation}
    \ce{H2} + \ce{O} \rightarrow \ce{OH} + \ce{H}
\end{equation}  
followed by
\begin{align}  
\ce{OH} + \ce{H2} -> \ce{H2O} + \ce{H} \, ,
\end{align} which have activation energies of $3150\,\rm K$ and $1740\,\rm K$, respectively \citep{McElroy2013}. Water is destroyed by photo-dissociation. Very close to the inner rim, stellar UV photons and X-rays penetrate the disk radially, which inhibits water formation. Once the radial column densities are large enough to shield the direct irradiation, the water concentration reaches its maximum value of about $10^{-4}$. Vertical integration down to the height where the dust becomes optically thick results in an about constant \ce{H2O} column density as function of radius in the selected disk setup.  The water line emitting disk surface ends quite abruptly where the gas temperature drops below about $200\,\rm K$ and the neutral-neutral reactions become inefficient.

The radial behaviour of temperature and column density overlaps well between ProDiMo and DuCKLinG from about $0.2\,\rm au$ outwards. Few water lines emit significantly further out (outermost emission from $6.6\,\rm au$) or closer to the star (innermost emission from $0.1\,\rm au$) in ProDiMo. The column densities retrieved by DuCKLinG fall also within the range of column densities of \ce{H2O} in the ProDiMo model. The column densities for \ce{H2O} in ProDiMo range from $4\times 10^{17} \mathrm{cm^{-2}}$ to $5\times 10^{20} \mathrm{cm^{-2}}
$, which is larger than the range retrieved by DuCKLinG ($3\times 10^{18} \mathrm{cm^{-2}}$ to $3\times 10^{20} \mathrm{cm^{-2}}$). The innermost very hot region (up to $1000\,\rm K$) of water emission in ProDiMo is not reproduced by DuCKLinG (ranging only from about $200\,\rm K$ to $400\,\rm K$). Analysing the cumulative integrated line flux (upper panel of Fig.~\ref{fig:metadata_prodimo}), we find that the hot inner region ($\lesssim 0.13\,\rm au$) has an integrated line flux less than $3\,\%$ of the total flux ($2.3 \times 10^{-17} \,\rm W/m^2$ to $8.8 \times 10^{-16} \,\rm W/m^2$). While weak lines can provide crucial information about conditions and processes in disks \citep[e.g.][]{Banzatti2023b}, an automated fit of the full wavelength range naturally focuses on the strongest lines. For analysing weak lines in detail, a more focused approach is needed. Overall, there is good agreement between the emitting conditions in ProDiMo and the retrieved values from DuCKLinG. This makes us confident that the retrieved values by DuCKLinG are meaningful when fitting MIRI spectra.

\subsection{Application to GW\,Lup\label{sec:gwlup}}

We fitted the JWST/MIRI spectrum of GW\,Lup using DuCKLinG to determine the continuum and molecular emission properties, including their uncertainties. This spectrum was published and fitted previously by \cite{Grant2023}, using a continuum subtraction by hand followed by a sequence of $\chi^2$-fitting of OD slab models for single molecules as explained below.  We will compare these published results to the results that we obtained from our simultaneous Bayesian fit of all molecules and the continuum.

GW\,Lup is an M\,1.5 star \citep{Alcala2017} at a distance of $155\,\rm pc$ \citep{Gaia2020}. The DSHARP survey \citep{Andrews2018} found that the continuum emission shows a narrow ring at a radius of $85\,\rm au$ \citep{Dullemond2018}. Spectra with the Spitzer Space Telescope revealed emission of \ce{C2H2} \citep{Banzatti2020} and strong emission of \ce{^{12}CO2} but no water \citep{Pontoppidan2010,Salyk2011}.  

\cite{Grant2023} fitted the JWST/MIRI spectrum of GW\,Lup between $13.6\,\rm \mu m$ and $16.3\,\rm \mu m$ and detected in addition to the with Spitzer detected molecules \ce{^{13}CO2} for the first time, and \ce{H2O}, \ce{HCN}, and \ce{OH} for the first time in this object. The fit to the MIRI-spectrum was obtained by \cite{Grant2023} by a step-by-step approach. First, the continuum was subtracted from the spectrum selecting $8$ points of the spectrum by hand that likely show no line emission and a cubic spline interpolation. Then, the molecular emissions were fitted one by one, and subtracted, using the following sequence: \ce{H2O}, \ce{HCN}, \ce{C2H2}, \ce{^{12}CO2}, \ce{^{13}CO2}, and \ce{OH}. The emission spectrum of every single molecule was fitted using 0D slab-models of temperature $T$, column density $\Sigma_{\rm col}$, and the radius of a circular emitting area $R$. The fit of every molecular emission spectrum was obtained by $\chi^2$-minimization using a grid of slab models.

%In this section, we fit the MIRI spectrum of GW\,Lup to test the fit quality our model can achieve on MIRI observations and compare the retrieved parameters to previous studies.

We fitted the MIRI-spectrum between $13.6\,\rm \mu m$ and $16.3\,\rm \mu m$ with DuCKLinG. While this wavelength range encompasses no strong dust features and makes the determination of the dust composition difficult (see Appendix~\ref{sec:dust_gwlup}), it allows for a comparison of the retrieved molecular properties to \cite{Grant2023}. The stellar temperature \citep[$3630\,\rm K$,][]{Alcala2017,Andrews2018}, distance \citep[$155\,\rm pc$,][]{Gaia2020}, and inclination \citep[$39^\circ$,][]{Andrews2018} were fixed to their respective literature values during our fitting.
The stellar radius ($1.45\,\mathrm{R_\odot}$) was calculated using the Stefan-Boltzmann law with given stellar temperature and luminosity \citep[$0.33\,\rm L_\odot$,][]{Alcala2017,Andrews2018}. 

Our fit includes \ce{H2O}, \ce{CO2}, \ce{^{13}CO2}, \ce{HCN}, and \ce{C2H2}. The column density ratio of \ce{^{13}CO2}:\ce{^{12}CO2} is fixed to 1:70. A ratio that falls within the allowable range for GW\,Lup derived by \cite{Grant2023}. Therefore, only the values for \ce{CO2} are reported, which determine the \ce{^{12}CO2} and \ce{^{13}CO2} values. Contrary to \cite{Grant2023}, we did not include \ce{OH} in our analysis since non-LTE effects might play a significant role for this species. This decision does not significantly impact the remaining species, because \cite{Grant2023} found only weak \ce{OH} features that do not overlap with the strongest features of the other species. The free parameters and their prior distributions are listed in Table~\ref{tab:prior_gwlup} (the prior ranges are motivated in Sect.~\ref{sec:prodimo}). We note that this table only includes the non-linear parameters. Additional linear parameters next to the midplane and rim scaling factors are the emitting radii of every molecule and the abundances of the different dust species. The linear parameters are automatically determined (see Sect.~\ref{sec:fitting}). The dust available during the fitting are am Mg-Olivine, am Mg-Pyroxene, Silica, Forsterite, and Enstatite, all with sizes of $0.1\,\rm\mu m$, $1.5\,\rm\mu m$, $2.0\,\rm\mu m$, and $5.0\,\rm\mu m$ as introduced in Table~\ref{tab:dustrefs}.

\begin{table}[t]
    \caption{Prior parameters ranges for fitting GW\,Lup's MIRI spectrum. $\mathcal{U}(x,y)$ and $\mathcal{J}(x,y)$ denote uniform and log-uniform priors in the range from $x$ to $y$, respectively. The units of the parameters are given in Table~\ref{tab:model_parameters}. Further linear parameters are not listed here, see text.}
    \label{tab:prior_gwlup}
    \centering
    \vspace*{-2mm}
    \begin{tabular}{l|l|l|l}
\hline \hline & & & \\[-1.9ex] 
Parameter & Prior & Parameter & Prior \\ \hline  
  & & & \\[-1.9ex] 
$T_{\rm rim}$ & $\mathcal{U}(10,1600)$ &$q_{\rm emis}$ & $\mathcal{U}(-1,-0.1)$ \\ 
 & & & \\[-1.9ex] 
$T^{\rm sur}_{\rm min}$ & $\mathcal{U}(10,400)$ &$a_{\rm obs}$ & $\mathcal{J}(10^{-5},10^{-1})$ \\ 
 & & & \\[-1.9ex] 
$T^{\rm sur}_{\rm max}$ & $\mathcal{U}(300,1600)$ &$\Sigma_{\rm tmin}$\tablefootmark{(1)} & $\mathcal{J}(10^{14},10^{24})$ \\ 
  & & & \\[-1.9ex] 
$T^{\rm mid}_{\rm min}$ & $\mathcal{U}(10,400)$ &$\Sigma_{\rm tmax}$\tablefootmark{(1)} & $\mathcal{J}(10^{14},10^{24})$ \\ 
  & & & \\[-1.9ex] 
$T^{\rm mid}_{\rm max}$ & $\mathcal{U}(300,1600)$ &$T_{\rm max}$\tablefootmark{(1)} & $\mathcal{U}(25,1500)$ \\  
 & & & \\[-1.9ex] 
$q_{\rm mid}$ & $\mathcal{U}(-1,-0.1)$ &$T_{\rm min}$\tablefootmark{(1)} & $\mathcal{U}(25,1500)$ \\  
 & & & \\[-1.9ex] 
$q_{\rm sur}$ & $\mathcal{U}(-1,-0.1)$ & & \\
    \end{tabular}
    
        \tablefoot{\\
\tablefoottext{1}{Same prior used for \ce{H2O}, \ce{CO2}, \ce{HCN}, and \ce{C2H2}.}}
\end{table}

\begin{figure*}[t]
    \centering
    \includegraphics[width=0.83\linewidth]{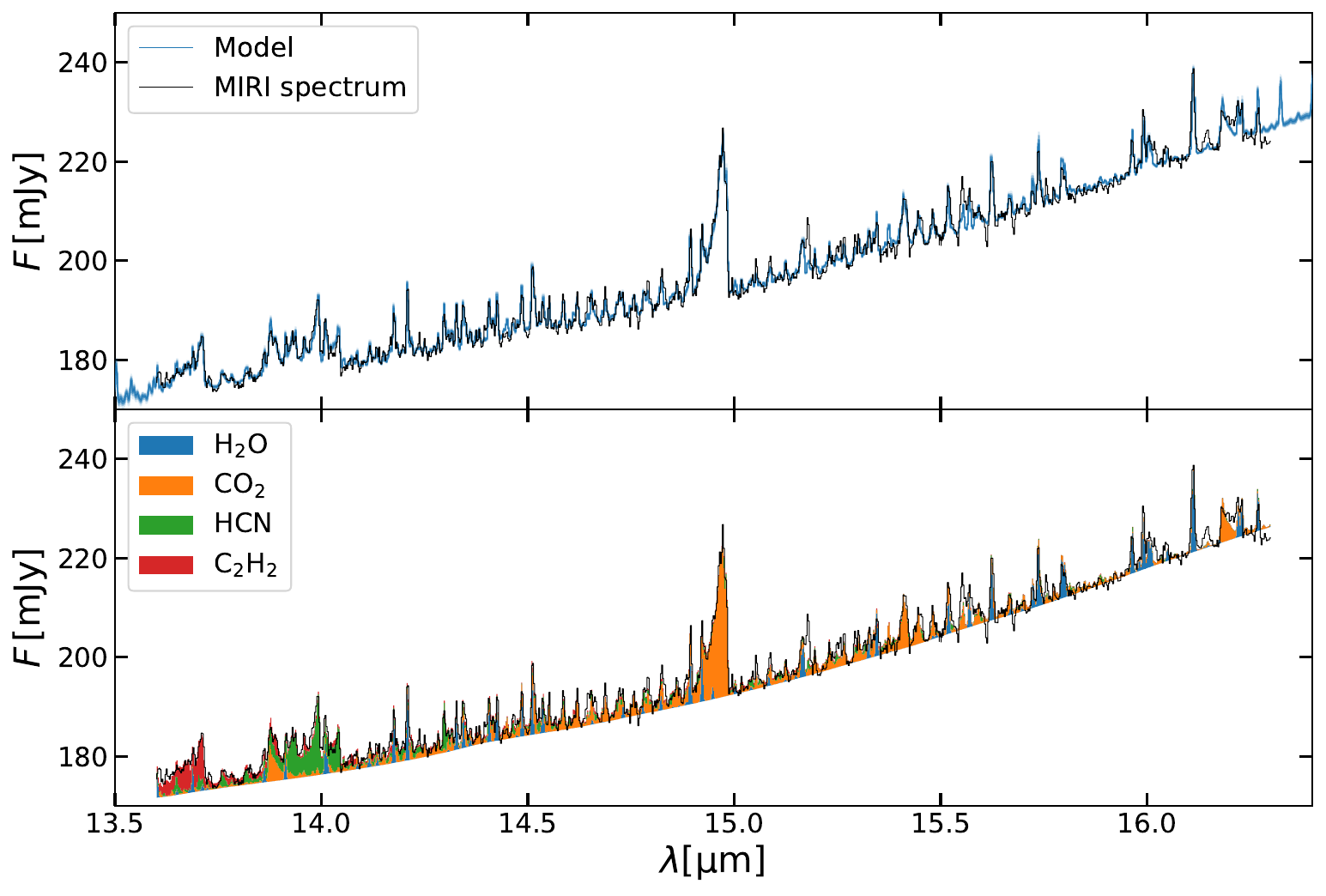}
    \caption{Upper panel: Spectral posterior distribution for a fit of the GW\,Lup MIRI spectrum. The black line represents the MIRI spectrum. The blue line shows the median fluxes from the models of the posterior distribution. The blue lines consist of contours that represent the $1\sigma$, $2\sigma$, and $3\sigma$ level of the fluxes. Lower panel: Median probability model. The black line represents the MIRI spectrum. The lower edge of the coloured region shows the continuum emission of the model, with the cumulative molecular emission shown in different colours.}
    \label{fig:gwlup_median_probable}
\end{figure*}

The model spectrum extracted from the posterior distribution of the parameters is shown in the upper panel of Fig.~\ref{fig:gwlup_median_probable}. The black line shows the MIRI spectrum of GW\,Lup in the fitted wavelength range. The median flux of the posterior of models is shown in blue. The $1\sigma$, $2\sigma$, and $3\sigma$ level of the fluxes are shown in decreasing intensities of blue. There is large overlap between the model fluxes and the observation. Additionally, the spread of model spectra is very small, which means that the Bayesian algorithm is converging to a narrow solution. We extract one representative model from the posterior distribution to examine the fit composition in more detail. This median probability model is the model for which all parameters are closest to their posterior median as outlined by \cite{Kaeufer2023}. While this model is not necessarily the model with the maximum likelihood, it is the closest to the median posterior values.

The different molecular contributions in this model are shown in the lower panel of Fig.~\ref{fig:gwlup_median_probable}. 
%There is a large overlap between the spectrum and the total flux of the model with 
The $1\sigma$ difference between our model and the observations is $1.21\,\rm mJy$ or $0.64\,\%$. This is comparable to the retrieved observational uncertainty of $0.72\,\%$. 
%It can be clearly seen how 
All main spectral features are well reproduced: the \ce{CO2} feature at about $15\,\rm\mu m$, the \ce{HCN} feature around $14\,\rm\mu m$, the \ce{C2H2} feature at about $13.7\,\rm\mu m$, and all water lines at wavelengths longer than $15.5\,\rm\mu m$. 
The few molecular lines that are present in the observation but not in the model (e.g.\ at about $15.2\,\rm \mu m$ or $16.1\,\rm \mu m$) hint to contributions of molecules that were not included in our model or lines of included molecules that are not included in the used data collection. 
%Additional evidence for this is provided by the fact that 
The slab-models by \cite{Grant2023} did not reproduce these lines as well, but focused their fits on narrow wavelength windows around the respective molecular features. If non-described emission becomes a substantial problem, it is possible with DuCKLinG to omit the respective wavelength ranges and derive a fit for the wavelength ranges that the model can describe.
%After examining the fit quality, we turn to the posterior distribution of the gas parameters. 

Our retrieved molecular properties with their posterior uncertainties are listed in Table~\ref{tab:gwlup_posterior} together with the corresponding values derived by \cite{Grant2023}. 
%This later fit included a single temperature and column density instead of ranges for both quantities.
The retrieved dust composition is discussed in Appendix~\ref{sec:dust_gwlup}.
However, we note that the small wavelength range of the fit does not include any strong dust features, and therefore the retrieved dust composition should be taken with a pinch of salt.

\begin{table}
    \caption{Posterior parameter values and uncertainties for selected parameters of the GW\,Lup fit. The third column shows the corresponding values found by \cite{Grant2023}. The notation $a(+b)$ means $a\times10^b$.}
    \centering
    \vspace{-2mm}
    \resizebox{!}{81mm}{
    \begin{tabular}{l l|l|l}
\hline \hline & & & \\[-1.9ex] 
\multicolumn{2}{l|}{Parameter} & Posterior & \cite{Grant2023} \\ \hline  
  & & & \\[-1.9ex] 
\multicolumn{2}{c|}{$T_{\rm rim}$} & $1210^{+270}_{-400}$ & \\ 
 & & & \\[-1.9ex] 
\multicolumn{2}{c|}{$T^{\rm sur}_{\rm min}$} & $240^{+110}_{-130}$ & \\ 
 & & & \\[-1.9ex] 
\multicolumn{2}{c|}{$T^{\rm sur}_{\rm max}$} & $1000^{+400}_{-400}$ & \\ 
 & & & \\[-1.9ex] 
\multicolumn{2}{c|}{$T^{\rm mid}_{\rm min}$} & $108^{+11}_{-17}$ & \\ 
 & & & \\[-1.9ex] 
\multicolumn{2}{c|}{$T^{\rm mid}_{\rm max}$} & $900^{+400}_{-400}$ & \\ 
 & & & \\[-1.9ex] 
\multicolumn{2}{c|}{$q_{\rm mid}$} & $-0.18^{+0.06}_{-0.09}$ & \\ 
 & & & \\[-1.9ex] 
\multicolumn{2}{c|}{$q_{\rm sur}$} & $-0.63^{+0.31}_{-0.25}$ & \\ 
 & & & \\[-1.9ex] 
\multicolumn{2}{c|}{$q_{\rm emis}$} & $-0.59^{+0.04}_{-0.06}$ & \\ 
 & & & \\[-1.9ex] 
\multicolumn{2}{c|}{$a_{\rm obs}$} & $0.00723^{+0.00015}_{-0.00015}$ & \\ 
 & & & \\[-1.9ex] 
\cline{1-2} 
&$r_{\rm eff}$ & $0.111^{+0.018}_{-0.008}$ &0.15\,au\\ 
 & & & \\[-1.9ex] 
&$t_{0.85}$ & $490^{+70}_{-40}$ &\multirow{2}*{625\,K}\\ 
 & & & \\[-1.9ex] 
\ce{H2O}&$t_{0.15}$ & $810^{+180}_{-130}$ &\\ 
 & & & \\[-1.9ex] 
&$\Sigma_{0.85}$ & $6^{+12}_{-5}(+18)$ &\multirow{2}*{3.2 (+18)\,$\rm cm^{-2}$}\\ 
 & & & \\[-1.9ex] 
&$\Sigma_{0.15}$ & $3.5^{+8}_{-2.1}(+18)$ &\\ 
 & & & \\[-1.9ex] 
\cline{1-2} 
&$r_{\rm eff}$ & $0.111^{+0.008}_{-0.008}$ &0.11\,au\\ 
 & & & \\[-1.9ex] 
&$t_{0.85}$ & $242^{+11}_{-10}$ &\multirow{2}*{400\,K}\\ 
 & & & \\[-1.9ex] 
\ce{CO2}&$t_{0.15}$ & $750^{+50}_{-60}$ &\\ 
 & & & \\[-1.9ex] 
&$\Sigma_{0.85}$ & $3.3^{+0.7}_{-0.5}(+18)$ &\multirow{2}*{2.2 (+18)\,$\rm cm^{-2}$}\\ 
 & & & \\[-1.9ex] 
&$\Sigma_{0.15}$ & $9.7^{+4.0}_{-3.1}(+17)$ &\\ 
 & & & \\[-1.9ex] 
\cline{1-2} 
&$r_{\rm eff}$ & $0.0496^{+0.0030}_{-0.0030}$ &0.06\,au\\ 
 & & & \\[-1.9ex] 
&$t_{0.85}$ & $570^{+60}_{-60}$ &\multirow{2}*{875\,K}\\ 
 & & & \\[-1.9ex] 
\ce{HCN}&$t_{0.15}$ & $750^{+90}_{-70}$ &\\ 
 & & & \\[-1.9ex] 
&$\Sigma_{0.85}$ & $9.7^{+5.0}_{-3.4}(+17)$ &\multirow{2}*{4.6 (+17)\,$\rm cm^{-2}$}\\ 
 & & & \\[-1.9ex] 
&$\Sigma_{0.15}$ & $2.3^{+1.4}_{-0.7}(+17)$ &\\ 
 & & & \\[-1.9ex] 
\cline{1-2} 
&$r_{\rm eff}$ & $0.029^{+0.009}_{-0.005}$ &0.05\,au\\ 
 & & & \\[-1.9ex] 
&$t_{0.85}$ & $540^{+130}_{-110}$ &\multirow{2}*{500\,K}\\ 
 & & & \\[-1.9ex] 
\ce{C2H2}&$t_{0.15}$ & $760^{+190}_{-140}$ &\\ 
 & & & \\[-1.9ex] 
&$\Sigma_{0.85}$ & $3.4^{+7}_{-2.5}(+17)$ &\multirow{2}*{4.6 (+17)\,$\rm cm^{-2}$}\\ 
 & & & \\[-1.9ex] 
&$\Sigma_{0.15}$ & $3.2^{+12}_{-2.3}(+17)$ &\\ 
 & & & \\[-1.9ex] 
\cline{1-2} 
 & & & \\[-1.9ex] 
\ce{H2O}&$\mathcal{N}_{\rm tot}$ & $7.1^{+3.3}_{-1.6}(+43)$ &5 (+43)\\ 
 & & & \\[-1.9ex] 
\cline{1-2} 
 & & & \\[-1.9ex] 
\ce{CO2}&$\mathcal{N}_{\rm tot}$ & $3.7^{+1.1}_{-0.8}(+43)$ &1.7 (+43)\\ 
 & & & \\[-1.9ex] 
\cline{1-2} 
 & & & \\[-1.9ex] 
\ce{HCN}&$\mathcal{N}_{\rm tot}$ & $1.6^{+0.7}_{-0.4}(+42)$ &1.2 (+42)\\ 
 & & & \\[-1.9ex] 
\cline{1-2} 
 & & & \\[-1.9ex] 
\ce{C2H2}&$\mathcal{N}_{\rm tot}$ & $3.4^{+4}_{-1.7}(+41)$ &9.3 (+41)\\ 
[+0.3ex]\hline
    \end{tabular}}
    \label{tab:gwlup_posterior}
\end{table}

\begin{figure}
    \centering
    \includegraphics[width=\linewidth]{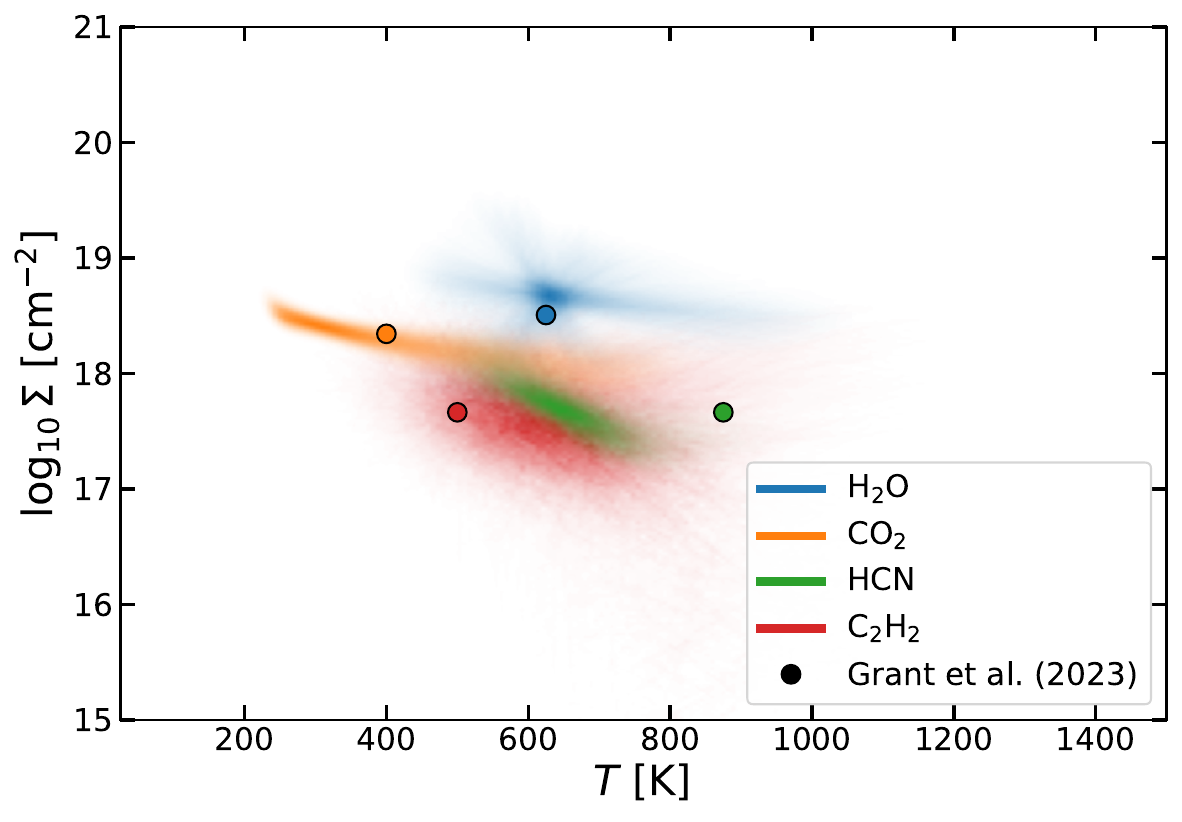}\\[-2mm]
    \caption{Molecular emission properties of GW\,Lup. The different coloured areas denote the parameter areas in which the respective molecule is significantly (as introduced in Sec.~\ref{sec:prodimo}) emitting in the posterior of models. The points show the retrieved parameter values by \cite{Grant2023} for the same molecules.}
    \label{fig:gwlup_mols}
\end{figure}

Our posterior distributions of the retrieved molecular column densities and emission temperatures are furthermore visualised in Fig.~\ref{fig:gwlup_mols}, keeping in mind that we use power laws for these quantities as function of radius for every molecule. 
%shows under which temperature and column densities the different molecules are emitting within the posterior. 
The shaded areas colour-code how many of the posterior models emit at these conditions (between $T_{0.15}$ and $T_{0.85}$ as introduced in Sect.~\ref{sec:prodimo}). 

The column density ranges from our fit always include the values derived by \cite{Grant2023}, at least within the $1\sigma$ ranges of its limits (Fig.~\ref{fig:gwlup_mols}).  The temperature ranges for \ce{H2O} and \ce{CO2} include the temperature values found by \cite{Grant2023}. 
%This can be seen in Fig.~\ref{fig:gwlup_mols} as well. 
%For both molecules, the conditions from \cite{Grant2023} fall close to the regions in which most posterior models emit. 
Interestingly, the water emissions in our model seem to originate from varying power laws that intersect at a point close to the value of \cite{Grant2023} (see Fig.~\ref{fig:gwlup_mols}). Therefore, the water emission from this area seems to be of great importance to reproduce the observed features.

For \ce{C2H2}, the previously found value of $500\,\rm K$ falls within the $1\sigma$-uncertainty of the lower $T$-limit found by this study ($540\,\rm K$), but our models also suggest contributions from significantly higher temperatures. Figure~\ref{fig:gwlup_mols} shows how our posterior emitting conditions for \ce{C2H2} overlap with the best fit value derived by \cite{Grant2023}.

 For \ce{HCN}, the previously found value ($875\,\rm K$) is significantly higher than the upper limit of our retrieved temperature interval up to $750\,\rm K$, even including the $1\sigma$ uncertainty of this limit. This can also be seen in Fig.~\ref{fig:gwlup_mols}, where the value from \cite{Grant2023} falls outside of the posterior distribution of the emission properties for HCN.  This difference might originate from our use of temperature and column density ranges, the different treatment of the continuum, or the fitting procedure in both studies. In Sect.~\ref{sec:bayes_gwlup} we show that the change from single values to ranges for the temperature and column density is not responsible for the retrieved differences and elaborate further on the potential reasons.
%The manual continuum determination and subtraction by \cite{Grant2023} results at short wavelength in a higher continuum flux, which might explain that the differences are only present for \ce{HCN} and especially \ce{C2H2} which have their strongest features at these wavelengths.

Comparing the emitting areas, we find that the values retrieved in our study are slightly smaller than the values found by \cite{Grant2023} (except for \ce{CO2} which has identical values in both approaches). Due to our inclusion of an inclination, we expect the values for $R_{\rm eff}$ to be slightly larger (factor of $(\cos 39^\circ)^{-1/2} \simeq 1.13$) than the ones found by \cite{Grant2023}. We explain this by the column density ranges that generally reach higher densities in our model than the single values, which in turn requires a smaller emitting area to reproduce the same flux. This can be tested by calculating the total number of molecules contributing to the total flux per molecule. 
The total number of molecules ($\mathcal{N}_{\rm tot}^{\rm mol}$) is calculated as the integral of column densities $\Sigma_{\rm mol}$ over the annuli from the inner ($R_{0.15}$) to the outer ($R_{0.85}$) radius with significant emission: 
\begin{align}
    \mathcal{N}_{\rm tot}^{\rm mol} = \int\displaylimits_{R_{0.15}^{\rm mol}}^{R_{0.85}^{\rm nol}} 2\pi r\,\Sigma_{\rm mol}(r)\,dr \label{eq:nuber_mol}
\end{align}
 
 These total numbers of molecules are compared at the end of Table~\ref{tab:gwlup_posterior}. We expect, similarly to the radius, that the values of this study are increased by a small factor ($(\cos 39^\circ)^{-1} \simeq 1.29$) compared to the previously derived results. The values from \cite{Grant2023} fall within or close to the $1\sigma$ uncertainties for \ce{H2O} and \ce{HCN}. \ce{C2H2} has the largest uncertainty. We will show in Sect.~\ref{sec:bayes_gwlup} that the simultaneous fitting of all molecules enables us to account for the contributions of other molecules at the wavelength range of the \ce{C2H2} feature. This decreases the amount of \ce{C2H2} needed to reproduce the spectrum.

The differences for \ce{CO2} can be explained by the different treatment of \ce{^{13}CO2} in our modelling approach. To allow for line overlap between isopologues, we fixed the ratio of  \ce{^{13}CO2} to \ce{^{12}CO2} to 1:70, while \cite{Grant2023} treats them as independent species, allowing for other ratios and differences in temperature. However, \cite{Grant2023} finds the literature value of the ratio to be within the allowed range and the retrieved temperature of \ce{^{13}CO2} is only $75\,\rm K$ lower than the temperature of \ce{^{12}CO2}. This is smaller than the temperature range retrieved by DuCKLinG and on a similar scale as the temperature error bars on the higher temperature limit. This makes us confident that constraining the fit to a fixed \ce{^{13}CO2} to \ce{^{12}CO2} ratio is justified by the benefit of line overlap. The total number of \ce{^{13}CO2} molecules we retrieve is $5.3^{+1.6}_{-1.2}(+41)$ which is close to the value of $9.3(+41)$ found by \cite{Grant2023}. Therefore, our slightly larger total number of \ce{^{12}CO2} molecules is needed to produce a similarly strong \ce{^{13}CO2} feature.

While there is a large overlap between the molecular properties retrieved by us in this study and by \cite{Grant2023}, we note that the advantage of our method is (i) the inclusion of an automated fitting of the continuum and (ii) the simultaneous Bayesian fitting of all parameters. It seems that for this particular object, the manual continuum subtraction by \cite{Grant2023} worked quite well. However, this procedure becomes more challenging when optically thick molecular line emission creates a quasi-continuum \citep{Tabone2023}. Additionally, the considerable difference (e.g. total number of \ce{C2H2} molecules) might be due to propagating errors with the iterating molecular fitting process. The Bayesian fitting of all molecules does circumvent this, especially if the molecular features are overlapping.

Besides the retrieval of the molecular emission properties, we simultaneously extract some dust properties with our fitting routine, see upper half of Table~\ref{tab:gwlup_posterior}. The temperature parameters (especially the maximum temperatures) are poorly constrained due to the small wavelength range ($13.6\,\mu m - 16.3\,\mu m$) of the fitted observation. Therefore, the provided uncertainties are dominated by the provided priors. The power law exponent of the gas emission ($q_{\rm emis}$) is relatively well constrained and falls within the poorly constrained range of the dust surface layer exponent ($q_{\rm sur}$). Similar values might hint towards a common layer, located at similar heights over the disk midplane, from which both the optically thin dust emission and the molecular emission lines are emitted.

The last non-molecular parameter listed in Table~\ref{tab:gwlup_posterior} is $a_{\rm obs}$ which denotes the observational uncertainty relative to the observed flux (see Eq.~\ref{eq:flux_uncertainty}). The retrieved uncertainty of $0.723\,\%$ results in an average $\sigma_{\rm obs}$ of $1.43\,\rm mJy$, which is larger than the value of $0.44\,\rm mJy$ determined by \cite{Grant2023} based on the noise of a line free region from $15.9\,\rm \mu m$ to $15.94\,\rm \mu m$. We note that the retrieved $a_{\rm obs}$ is not model-independent, since it denotes the uncertainty that best explains the differences between model and observation. Therefore, it can be seen as an upper estimate for the observational uncertainty of the fitted spectrum which makes it consistent with the previously determined value.

\section{Discussion\label{sec:discussion}}

\subsection{GW\,Lup model complexity\label{sec:bayes_gwlup}}

\begin{table}[t]
    \centering
    \caption{Bayes factors of different fits of GW\,Lup. The original fit from Sect.~\ref{sec:gwlup} is compared to fits with less complexity. 
    Every fit considers one molecule to be emitted from a single column density or additionally from a single temperature.}
    \label{tab:bayes_gwlup}
    \vspace*{-2mm}
\begin{tabular}{l|p{1cm}|p{1cm}|l|l|l}
\hline \hline
&&&&&\\[-1.9ex]
mol\tablefootmark{(1)} & $\Sigma$ range & $T$ range & $\ln B$\tablefootmark{(2)} & Pref\tablefootmark{(3)} & Evidence\tablefootmark{(4)} \\ \hline
&&&&&\\[-1.9ex]
 \ce{H2O} & no & yes & -0.00 & no & none \\ 
 \ce{H2O} & no & no & -2.42 & no & weak \\ 
 \ce{CO2} & no & yes & -1.61 & no & weak \\ 
 \ce{CO2} & no & no & -21.55 & no & very strong \\ 
 \ce{HCN} & no & yes & 0.26 & yes & none \\ 
 \ce{HCN} & no & no & 0.86 & yes & none \\ 
 \ce{C2H2} & no & yes & 0.17 & yes & none \\ 
 \ce{C2H2} & no & no & -0.21 & no & none \\ 
 all & no & no & -23.16 & no & very strong \\ 
 \hline
\end{tabular}
    \tablefoot{\\
\tablefoottext{1}{Molecule that is sampled differently compared to the original fit.}\\
\tablefoottext{2}{Logarithm of the Bayes factor between this fit and the original one.}\\
\tablefoottext{3}{Is this model preferred over the original one?}\\
\tablefoottext{4}{Interpretation of $B$ based on \cite{Trotta2008}.}
}
\end{table}

 Our Bayesian analysis allows to quantify the benefits of using simple vs.\ more complex models. In the DuCKLinG models, the molecules are allowed to emit from radial ranges of column densities and emission temperatures, but we can also enforce single values of these quantities. In the most simplified case, the model falls back to simultaneous 0D slab-model fits on top of the dust continuum.

Table~\ref{tab:bayes_gwlup} compares different models of reduced complexity, in application to the GW\,Lup spectrum, to the full model as discussed in Sect.~\ref{sec:gwlup}. The comparison is done using the Bayes factor ($B$), which compares the evidence ($p(d|M_1)$) of the observed data ($d$) for one model $M_1$ to the evidence ($p(d|M_2)$) for another model $M_2$:
 \begin{align}
     B= \frac{p(d|M_1)}{p(d|M_2)}
 \end{align}
$\ln B<1$, $1<\ln B<2.5$, $2.5<\ln B<5$, $5<\ln B<11$, and $11<\ln B$ correspond to no evidence, weak evidence, moderate evidence, strong evidence, and very strong evidence, respectively, for a variant model ($M_1$) over the full model ($M_2$) \citep{Trotta2008}. The sign of $B$ indicates which model is preferred, with negative values meaning that the full model is preferred.

 Table~\ref{tab:bayes_gwlup} lists these Bayes factors for altogether 9 models, using either a single column density value or single values for both the column density and the emission temperature, for one selected molecule as indicated. All other molecules are still fitted using ranges for both column densities and emission temperatures.
 %Additionally, $4$ models reduce the temperature range of a particular molecule to a single value. 
 %Therefore, this single column density and temperature version of the model corresponds to a single slab model on top of the dust continuum. 
 Additionally, one model labelled with  `all' uses 0D slab-models (single values for column density and temperature) for all molecules.

We find no evidence that varying column densities and emission temperatures are required to fit the observations of \ce{HCN} and \ce{C2H2} for the case of GW\,Lup, which hints that the emission regions for these molecules do probably not show a large diversity in conditions.
For \ce{CO2} there is weak evidence that a column density range is needed ($\ln B\!=\!-1.61$) and very strong evidence against a single column density and temperature description ($\ln B\!=\!-21.55$). Therefore, it seems that \ce{CO2} is emitted from a region that varies in column density and temperature. Similarly, we find weak evidence ($\ln B\!=\!-2.42$) that varying emission temperatures are required for \ce{H2O}, given the observations.

For the model that used 0D slab models for all molecules, there is very strong evidence (logarithm of the Bayes factor of $23.16$) that it cannot reproduce the observations as well as our full model. This is largely driven by the worse fitting of \ce{H2O} and \ce{CO2} by the 0D slabs, since the Bayes factors for the reduced model of only \ce{HCN} and \ce{C2H2} are not significant. 

\begin{figure}
    \centering
    \includegraphics[width=\linewidth]{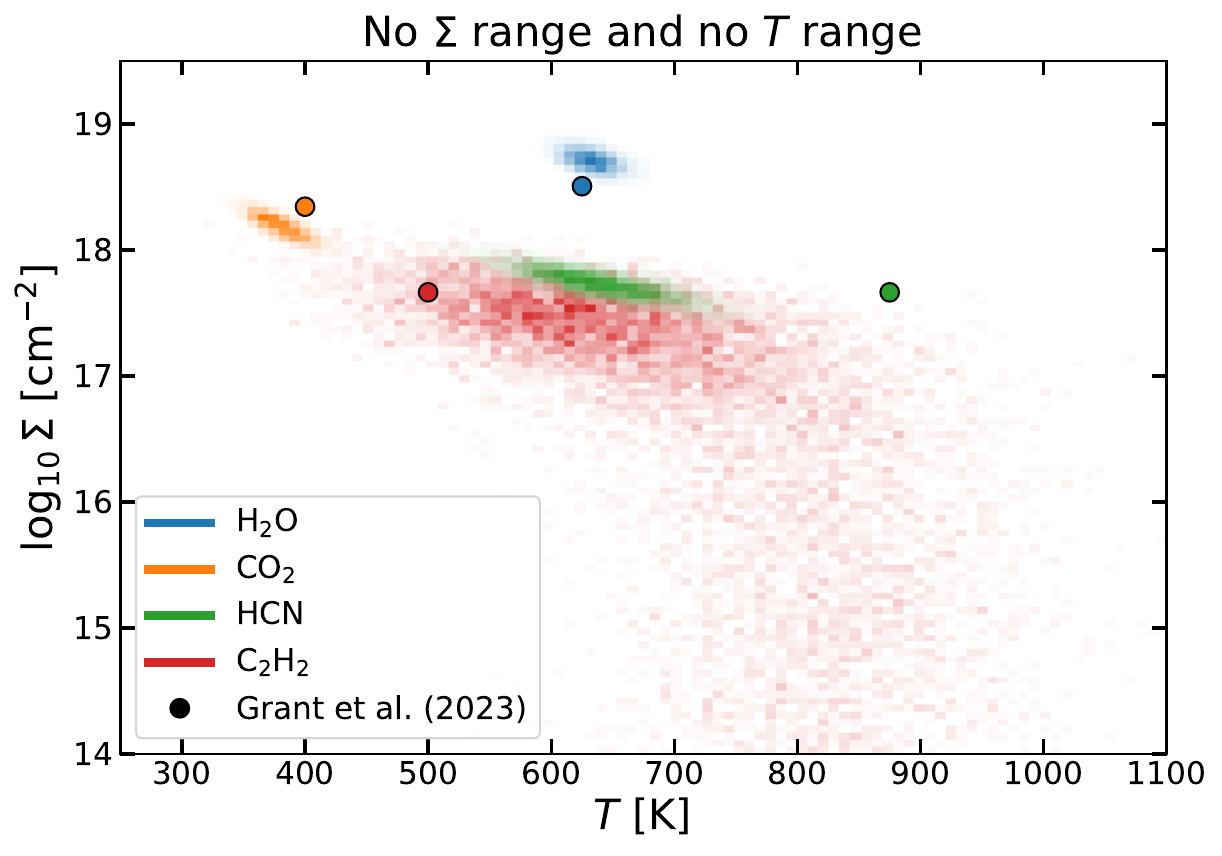}\\[-2mm]
    \caption{Molecular emission properties of GW\,Lup for the 0D slab retrieval. The different coloured areas denote the parameter area under which the respective molecule is emitting in the posterior of models. The points show the retrieved parameter values by \cite{Grant2023} for the same molecules.}
    \label{fig:gwlup_mols_0Dslab}
\end{figure}

However, this model allows for a comparison to the fitting results from \cite{Grant2023} with identical complexities for the molecular emission.
Fig.~\ref{fig:gwlup_mols_0Dslab} shows the emission conditions retrieved by the model using 0D slabs for all molecules. The shared areas result from the variation of temperature and column density in the posterior of models.

For both water (temperature: $633^{+16}_{-15}\,\rm K$; column density: $5.0^{+0.9}_{-0.8}(+18)\,\rm cm^{-2}$) and \ce{CO2} (temperature: $380^{+18}_{-17}\,\rm K$; column density: $1.54^{+0.32}_{-0.32}(+18)\,\rm cm^{-2}$) the posterior is well constrained to values close to the ones extracted by \cite{Grant2023} (see Table~\ref{tab:gwlup_posterior} for the values). This supports the interpretation that the observed water features require emission conditions close to the values from \cite{Grant2023}.

The conditions of \ce{C2H2} are rather poorly constrained with the temperature ($690^{+140}_{-120}\,\rm K$) having error bars larger than $100\,\rm K$ and the column density having a lower $1\sigma$ limit of $3.3(+15)\,\rm cm^{-2}$ nearly two orders of magnitudes lower than the median value $1.5(+17)\,\rm cm^{-2}$. The large uncertainty in column density is most likely due to the gas being optically thin, which makes the emitting area and column density completely degenerate. This degeneracy of \ce{C2H2} is also found by \cite{Grant2023}.

Most striking are the retrieved differences for \ce{HCN}. While the column density retrieved by \cite{Grant2023} falls within the by DuCKLinG retrieved range ($5.3^{+1.5}_{-1.2}(+17)\,\rm cm^{-2}$), the temperature of $875\,\rm K$ shows large disagreement ($640^{+50}_{-40}\,\rm K$). A similar difference is seen in the full retrieval (Sect.~\ref{sec:gwlup}). Therefore, we conclude that the difference is not due to the different treatment of the continuum and fitting procedure instead of the used model complexity. It can be seen that the \ce{HCN} feature at about $14\,\mu m$ (Fig.~\ref{fig:gwlup_median_probable}) shows a non-negligible flux from \ce{CO2} over the full wavelength range as well. While the Bayesian analysis employed in this study optimises the \ce{CO2} and \ce{HCN} conditions at the same time, the iterative fitting procedure by \cite{Grant2023} fits \ce{HCN} before subtracting the flux from \ce{CO2}. 
Therefore, we speculate that the iterative fitting is the origin of the discrepancy in retrieved temperatures for \ce{HCN} between this study and \cite{Grant2023}. This underlines the need for Bayesian analysis with a model that describes the molecular emission by all molecules and the dust emission at the same time to interpret JWST/MIRI spectra.

\section{Summary and conclusion\label{sec:summary}}

In this paper, we introduce DuCKLinG, a Python-based computer code to simultaneously model the dust continuum and the molecular emission properties of protoplanetary disks. The model is a superposition of optically thick and thin dust emission based on the dust opacity models by \cite{Juhasz2009,Juhasz2010} and slab models describing the molecular emissions from a handful of molecules in LTE. While previous studies used single values for column density and emission temperature for each molecule, we allow for radial powerlaw distributions of column densities and emission temperatures for each molecule to describe the spectra observed with JWST/MIRI.
The model is very flexible and applicable in many cases (e.g. different stellar types, disk structures, and inclinations). Possible limitations are the exclusion of non-LTE effects, which \cite{Banzatti2023a} show to be relevant for water, especially at short wavelength ($<9\,\rm \mu m$), and the lack of absorption lines, due to the independent treatment of dust and gas. This might be worth exploring in future projects.
Our model spectra can be compared to observations employing a full Bayesian analysis, since the code is speed-optimized and requires only about $6$ CPU milliseconds to generate one model spectrum. This allows among other things for an automated analysis of large samples and analysis on the preference of different complexities (e.g. which molecules are emitting under diverse or homogenous conditions).
The main conclusions of this paper are:

\begin{itemize}
    \item Determining linear parameters with NNLS instead of Bayesian sampling decreases the computational time of the retrieval by a factor of about $80$ for a mock observation and does not significantly change the retrieved median posterior values.
    \item The model can reproduce a mock observation by ProDiMo with a $1\sigma$ deviation of only $2.5\,\rm mJy$. The retrieved emission conditions describing the \ce{H2O} emission correspond well with emitting conditions in complex thermo-chemical codes like ProDiMo.
    \item The model can successfully reproduce the MIRI spectrum of GW\,Lup without the need for a continuum subtraction.
    \item The retrieved molecular conditions for GW\,Lup overlap well with values extracted from slab models by \cite{Grant2023}. Temperature differences for \ce{HCN} are attributed to the simultaneous fitting procedure compared to iterative fitting and not to the introduction of temperature and column density ranges.
    \item We conclude that \ce{H2O} and \ce{CO2} are emitting from a radially extended region of GW\,Lup that varies significantly in temperature but at least for \ce{H2O} not in column density. The emission of \ce{HCN} and \ce{C2H2} originates from a narrow region that does not show a large variety of emitting conditions (column density and temperature).
    \item We show that it is possible to retrieve the optically thin dust composition at the example of GW\,Lup. However, we note that the selected wavelength range does not allow for reliable dust composition constraints and should be taken as motivation for future studies.
\end{itemize}

\begin{acknowledgements}
      We acknowledge funding from the European Union H2020-MSCA-ITN-2019 under grant agreement no. 860470 (CHAMELEON).
\end{acknowledgements}
%\bibpunct{(}{)}{;}{a}{}{,} % to follow the A&A style
\bibliographystyle{aa} % style aa.bst
\bibliography{lib.bib} % your references

\appendix
\section{Derivation of model components\label{app:model_paras}}

In this section, the integral substitution from Eq.~(\ref{eq:integral_mid}) to
 Eq.~(\ref{eq:midplane}) is shown. This is equivalent to the translations of Eq.~(\ref{eq:integral_sur}) to Eq.~(\ref{eq:surface}) and Eq.~(\ref{eq:emis_radial}) to Eq.~(\ref{eq:emission}). Using the temperature power law (Eq.~\ref{eq:temp_powerlaw_mid}) it follows that
\begin{align}
    \frac{dr}{dT} = \frac{1}{q}\frac{R_{\rm min}^{\rm  mid}}{T_{\rm max}^{\rm  mid}} \left(\frac{T}{{{T_{\rm max}^{\rm  mid}}}}\right)^{1/q_{\rm mid}-1}
\end{align}
This means that,
\begin{align}
    r\,dr = \frac{1}{q} \frac{{\left(R_{\rm min}^{\rm  mid}\right)}^2}{T_{\rm max}^{\rm  mid}} \left(\frac{T}{{{T_{\rm max}^{\rm  mid}}}}\right)^{(2-q_{\rm mid})/q_{\rm mid}} dT,
\end{align}
using the expression $2/q_{\rm mid}-1=(2-q_{\rm mid})/q_{\rm mid}$. Therefore, $r\,dr$ can be substituted in Eq.~(\ref{eq:integral_mid}) which will result in Eq.~(\ref{eq:midplane}).

The surface layer uses the same relations with their component's respective quantities with the additional factor of $N_{d}^{j}\times \sigma_{\nu}^{j}$.
Analogously, Eq.~(\ref{eq:emis_radial}) uses the molecular intensities instead of the black bodies of Eq.~(\ref{eq:integral_mid}), but the derivation stays the same.

\section{Calculation of the dust mass\label{app:dust_mass}}

In this section, we derive the optically thin mass per dust species based on $C_{\rm sur}^{j}$. The total mass of a dust species $M_{j}$ is given by
\begin{align}
    M &= \pi \left[\left({R_{\rm max}^{\rm sur}}\right)^2-\left({R_{\rm min}^{\rm sur}}\right)^2\right] \, \Sigma_j \label{eq:mass_dust}
\end{align}
using the assumption that the column density $\Sigma_{j}$ is constant within the disk.
Using the relation between temperature and radius 
\begin{align}
    R_{\rm sur}^{\rm max} = R_{\rm sur}^{\min} \left(\frac{T_{\rm sur}^{\rm min}}{T_{\rm sur}^{\rm max}}\right)^{(1/q_{\rm sur})} \ ,
\end{align}
Eq.~(\ref{eq:mass_dust}) can be translated to
\begin{align}
    M &= \pi \left[\left({R_{\rm min}^{\rm sur}}\right)^2 \, \left( \left(\frac{T_{\rm sur}^{\rm min}}{T_{\rm sur}^{\rm max}}\right)^{(2/q_{\rm sur})}-1 \right)  \right] \, \Sigma_j. \label{eq:mass_dust_two}
\end{align}
The column number density $N_{d}^j$ can be translated to the mass column density using
\begin{align}
    \Sigma_{j}=N_{d}^j \, m_{j} ,
\end{align}
where $m_{j}$ is the mass per dust grain.
Therefore, Eq.~(\ref{eq:mass_dust_two}) translates to
\begin{align}
    M &= \pi \left[\left({R_{\rm min}^{\rm sur}}\right)^2 \, \left( \left(\frac{T_{\rm sur}^{\rm min}}{T_{\rm sur}^{\rm max}}\right)^{(2/q_{\rm sur})}-1 \right)  \right] \, N_{d}^j \, m_{j}. \label{eq:mass_dust_number_density}
\end{align}
The inner radius of the surface layer and the column number density can be expressed by $C_{\rm sur}^j$ as shown in Eq.~(\ref{eq:surface}). This results in   
\begin{align}
    M &= -\left( \left(\frac{T_{\rm sur}^{\rm min}}{T_{\rm sur}^{\rm max}}\right)^{(2/q_{\rm sur})}-1 \right) \,   \frac{ m_{j}\,d^2 q_{\rm sur} \left({T^{\rm max}_{\rm sur}}\right)^{(2/q_{\rm sur})} C_{\rm sur}^{j}}{2\,\cos{i}}  \label{eq:mass_dust_number_density_two}
\end{align}
which only contains variables that are determined during the fitting procedure.

\section{The dust composition of GW\,Lup \label{sec:dust_gwlup}}

%While analysing the dust composition is not the primary goal of this study, in this section we explore the retrieved dust composition. 

One advantage of DuCKLinG is the simultaneous fitting of dust and gas, which allows for conclusions about the dust mineralogy in protoplanetary disks. In this section, we examine the retrieved dust composition of GW\,Lup for the fit presented in Sect.~\ref{sec:gwlup}. However, we note that the fitted wavelength region was selected to compare the molecular results to \cite{Grant2023} and not optimised to retrieve well-constrained dust properties. The limited wavelength region ($2.7\rm\,\mu m$) lacks clear dust features, which complicates the retrieval of the dust composition. Therefore, the results presented here should be taken as a motivation for the possible application of DuCKLinG and not as an attempt to get reliable dust constraints. 

\begin{figure}[t]
    \centering
    \includegraphics[width=0.9\linewidth]{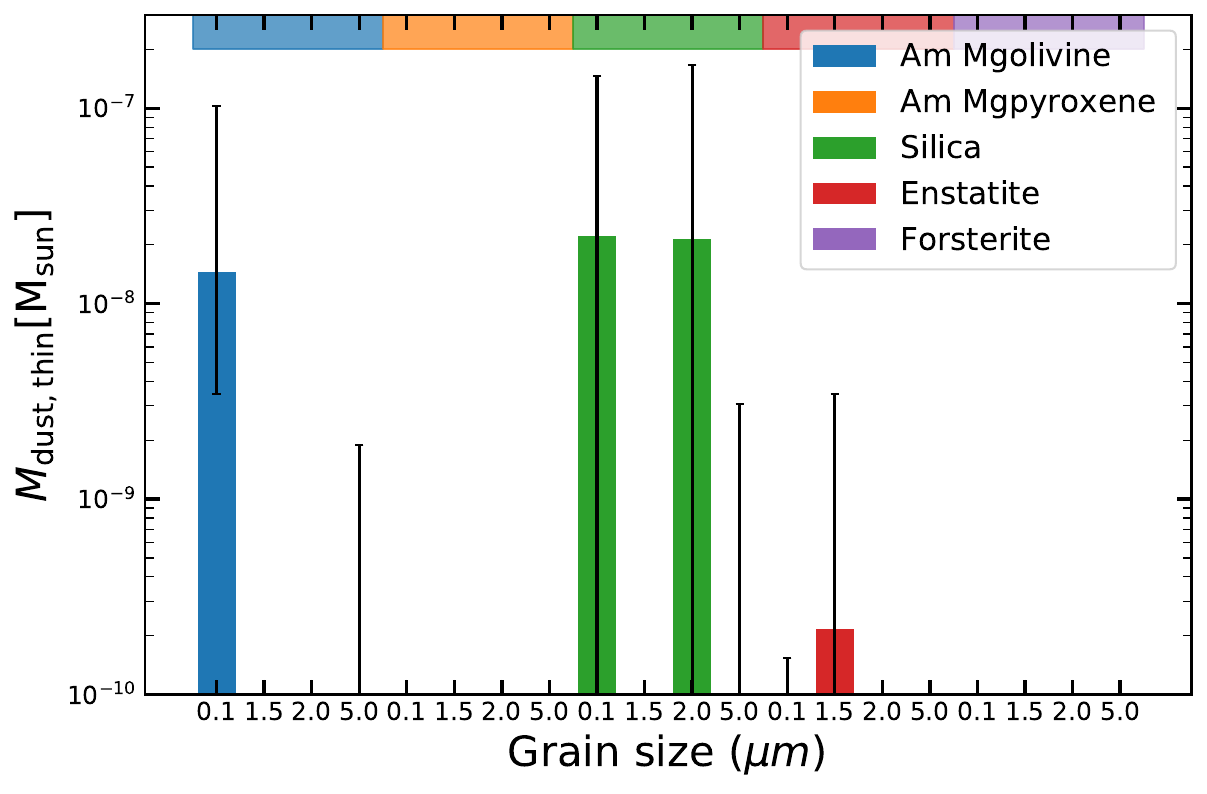}
    \caption{Histograms of the optically thin dust masses retrieved for GW\,Lup. The different dust species are colour-coded with the horizontal axis displaying the grain size. The histogram displays the median absolute optically thin dust mass ($M_{\rm dust, thin}$) of the respective species. It is accompanied by error bars that show the $1\sigma$ and $-1\sigma$ uncertainty.}
    \label{fig:gwlup_dust}
\end{figure}

The optically thin dust mass for all dust species included in the retrieval is shown in Fig.~\ref{fig:gwlup_dust}. 
A dust species is thereby defined as a material and a grain radius. 
Many dust species are not selected during the fitting procedure to reproduce the observation which is an effect of the NNLS fitting. The NNLS solver determines the dust scaling factors and therefore the dust masses to best reproduce the observation. A dust mass of $0\,M_\odot$ is thereby an option if the dust species does not improve the fit quality. Only four dust species are included in more than $25\,\%$ of the posterior models. These included species are Mg-Olivine grains of size $0.1\,\rm \mu m$, Silica grains of size $0.1\,\rm \mu m$ and $2.0\,\rm \mu m$ and Enstatite of size $1.5\,\rm \mu m$. All other species seem to be unimportant to reproduce the observation from $13.6\,\rm \mu m$ to $16.3\,\rm \mu m$.

Silica grains of sizes $0.1\,\rm \mu m$ and $2.0\,\rm \mu m$ have optically thin masses of $2.2\times10^{-8}\,M_\odot$ with an upper $1\sigma$ limit of $1.5\times10^{-7}\,M_\odot$, and $2.1\times 10^{-8}\,M_\odot$ with an upper $1\sigma$ limit of $1.7\times10^{-7}\,M_\odot$, respectively.
The olivine grains of size $0.1\,\rm \mu m$ have a mass of $1.4\times 10^{-8}\,M_\odot$. While the lower uncertainty of the dust masses for both silica grains is $0 \, M_\odot$, the lower limit of olivine is $3.4\times 10^{-9}\, M_\odot$. 
Additionally, enstatite grains of size $1.5\,\rm \mu m$ are included in the models of the posterior with a median mass of $8.9\times10^{-9}\,M_\odot$.

\begin{figure}
    \centering
    \includegraphics[width=0.9\linewidth]{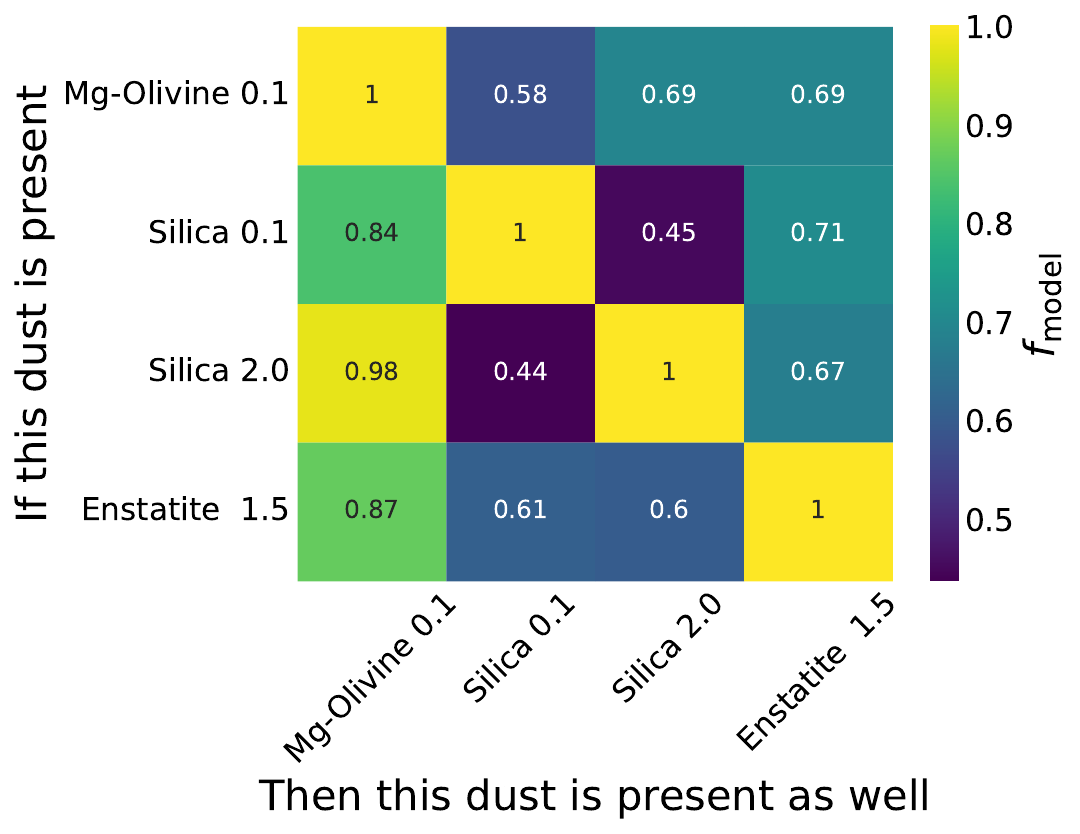}
    \caption{This heatmap shows which dust species and sizes are simultaneously used to describe the MIRI spectrum of GW\,Lup. The plot should be read row by row. One row indicated all models of the posterior that include the dust species named on the y-axis. The single entries show the faction of models ($f_{\rm model}$) that use the dust species on the x-axis at the same time.}
    \label{fig:dust_heatmap}
\end{figure}

While Fig.~\ref{fig:gwlup_dust} shows the retrieved dust masses for all species, it does not provide information about which dust species are used at the same time. Some models of the posterior might use one combination, while others use a different one with a similar effect. Therefore, Fig.~\ref{fig:dust_heatmap} indicates which dust species are used together. Every row selects only the models of the posterior for which the dust species listed on the vertical axis was used in significant abundance (more than $10^{-12}\,\,M_\odot$). Focusing on these models the horizontal axis indicates the faction of models ($f_{\rm model}$) that include the listed species simultaneously. 

Focusing on the second and third rows, it becomes clear that the models including Silica grain of size $0.1\,\rm \mu m$ and $2.0\,\rm \mu m$ only include the other Silica size in $45\%$ and $44\%$ of the cases, respectively. This shows that these two grains are degenerate enough that the differences can be compensated by the remaining dust species.
The left-most column of the heatmap shows that Mg-Olivine of size $0.1\,\rm \mu m$ is used in the vast majority of all cases, no matter which other dust species is selected.

We note that the dust properties could be much better constrained when we had used the full MIRI wavelength range. Therefore, the results presented here should be taken as a motivation to try using DuCKLinG or DuCK (Dust Continuum Kit \sout{with Line emission from Gas}) to determine the dust mineralogy similar to Jang et al. (in prep).

\end{document}